# Generative artificial intelligence enhances creativity but reduces the diversity of novel content

**Anil R. Doshi**[1] and **Oliver P. Hauser**[2]

[1]Department of Strategy and Entrepreneurship, UCL School of Management, London, UK
[2]Department of Economics, University of Exeter, Exeter, UK
Correspondence: anil.doshi@ucl.ac.uk and o.hauser@exeter.ac.uk

**Abstract.** Creativity is core to being human. Generative artificial intelligence (GenAI) holds promise for humans to be more creative by offering new ideas, or less creative by anchoring on GenAI ideas. We study the causal impact of GenAI on the production of a creative output in an online experimental study where some writers are could obtain ideas for a story from a GenAI platform. Access to GenAI ideas causes an increase in the writer's creativity with stories being evaluated as better written and more enjoyable, especially among less creative writers. However, GenAI-enabled stories are more similar to each other than stories by humans alone. Our results have implications for researchers, policy-makers and practitioners interested in bolstering creativity, but point to potential downstream consequences from over-reliance.

**One sentence summary**: Generative artificial intelligence helps produce more creative stories—but they are also more similar to each other.

Creativity is fundamental to innovation and human expression through literature, art, and music (*1*). However, the emergence of generative artificial intelligence (GenAI) technologies is challenging several long-standing assumptions about the uniqueness and superiority of human-generated content (*2*). GenAI is able to create new content in text (e.g., ChatGPT), images (e.g., Midjourney), audio (e.g., Jukebox), and video (e.g., Pictory). While GenAI has previously been shown to enable joint AI-human storyline development (*3*), increase quality and efficiency of production of typical white-collar work (*4*), promote productivity in customer support relations (*5,6*), speed up programming tasks (*7*), and enhance persuasion messaging (*8*), little is known about GenAI's potential impact on a fundamental human behavior: the ability of humans to be creative.

In this paper, we study how GenAI affects participants' ability to produce creative content. The domain we focus on is literature, an "open-creativity" task. Open-creativity tasks are critical in much of human activity and the economy, but have been found to be difficult to change or improve using economic incentives (*9*). Rather than employing incentives, our paper focuses on GenAI as a potential influence that affects humans' ability to be creative on an open-creativity task.

There are at least two ways in which the availability of GenAI can affect creativity. On the one hand, GenAI may enhance: generated ideas from AI may be used as a "springboard" for the human mind, providing potential starting points that can result in a "tree structure" of different storylines (*3,10*). It can also offer multiple starting venues that help a human writer overcome "writer's block" and the fear of a blank page (*11*). If this is the case, we would expect GenAI to lead to more creative output generated by human writers.

Conversely, GenAI may hamper: by anchoring the writer to a specific idea, or starting point for a story, GenAI may restrict the variability of a writer's own ideas from the start, inhibiting open creativity. Moreover, the output offered by GenAI may be derivative and thus not provide a fertile ground for new and creative ideas. If this is the case, we would expect GenAI to lead to more similar stories and potentially less creative output generated by human writers. Note that these two pathways in which GenAI can affect human creativity may not be mutually exclusive: it is possible that GenAI enhances human's ability to be creative in some ways (e.g. novelty) but not in others (e.g. usefulness) (*12*).

This paper aims to answer these questions through a pre-registered, two-phase experimental online study (see Figure 1 for the experimental design and Methods below for details). In the first phase of our study, we recruited a group of *N*=293 participants ("writers") who are asked to write a short, eight sentence story. Participants are randomly assigned to one of three conditions: *Human only*, *Human with 1 GenAI idea*, and *Human with 5 GenAI ideas* (see Table S1 for balance across conditions). In our *Human only* baseline condition, writers are assigned the task with no mention of or access to GenAI. In the two GenAI conditions, we provide writers with the option to call upon a GenAI technology (OpenAI's GPT-4 model) to provide a three-sentence starting idea to inspire their own story writing. In one of the two GenAI conditions (*Human with 5 GenAI ideas*), writers can choose to receive up to five GenAI ideas, each providing a possibly different inspiration for their story. After completing their story, writers are asked to self-evaluate their story on novelty, usefulness, and several emotional characteristics (see Supplementary Information (SI) Section 1 for all study questions).

In the second phase, the stories composed by the writers are then evaluated by a separate group of *N*=600 participants ("evaluators") (see Table S2 for balance across conditions). Evaluators read six randomly selected stories without being informed about writers being randomly assigned to access GenAI in some conditions (or not). All stories are evaluated by multiple evaluators on novelty, usefulness, and several emotional characteristics. After disclosing to evaluators whether GenAI was used during the creative process (*13*), we ask evaluators to rate the extent to which ownership and hypothetical profits should be split between the writer and the AI (*14*). Finally, we elicit evaluators' general views on the extent to which they believe that the use of AI in producing creative output is ethical, how story ownership and hypothetical profits should be shared between AI creators and human creators, and how AI should be credited in the involvement of the creative output (*15,16*).

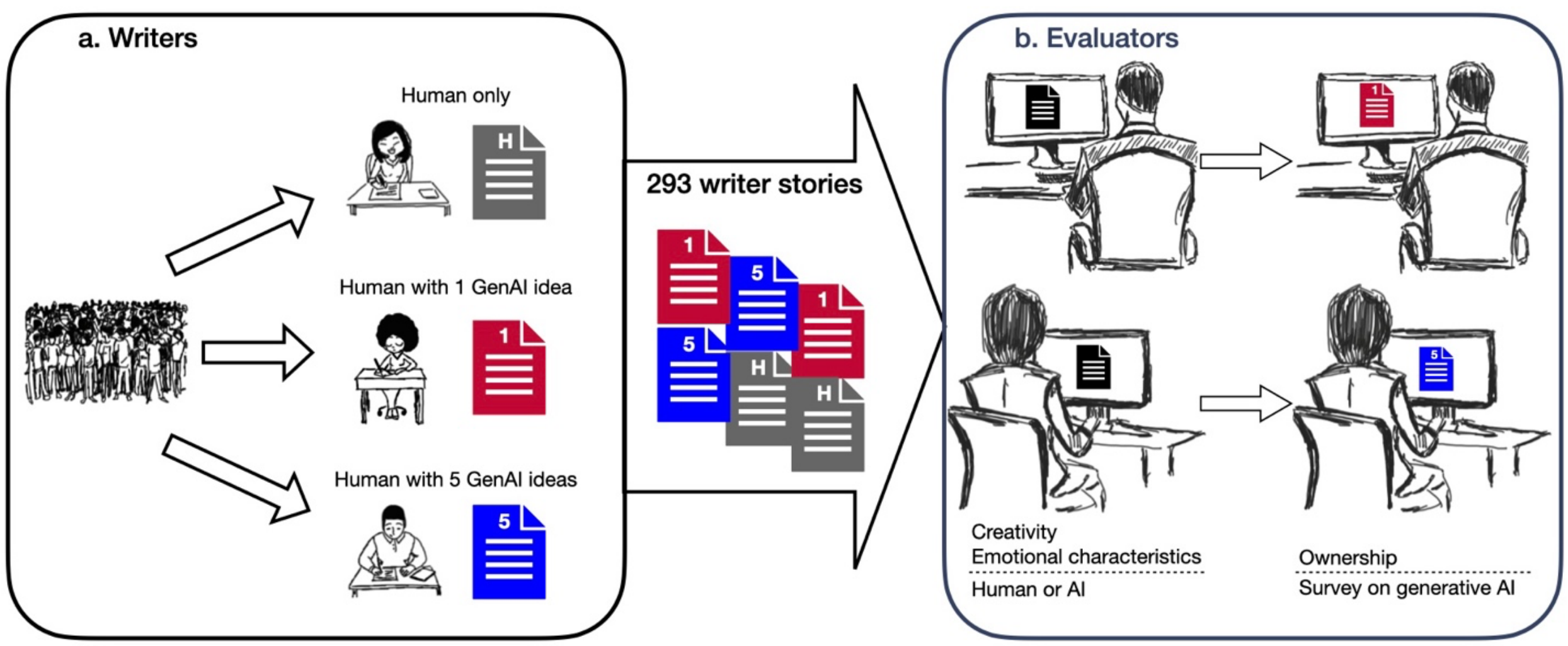

**Figure 1.** Visual representation of experimental design. **a**, Participants are recruited, provide consent to participate in the study, and complete the divergent association task (DAT)—a measure of an individual's inherent creativity (see Olson et al. (*17*))—before being randomly assigned to one of three experimental conditions: a *Human only* condition where the story was written with no GenAI assistance, a *Human with 1 GenAI idea* condition, and a *Human with 5 GenAI ideas* condition. A total of 293 stories are collected and then passed to evaluators. **b**, Evaluators provide ratings on six randomly assigned stories. The evaluators cycle through each story three times. First, prior to any information revelation, the evaluator assesses the creativity and emotional characteristics of the story. Second, the evaluator is asked to assess how likely the story was written by an AI versus a human. Third, the evaluator is told about whether the writer had access to and used GenAI and then provides responses about the ownership claim of the writer of each story. Evaluators then provide general responses to their views of GenAI.

## Results

**Baseline versus both GenAI combined.** We first estimate whether—relative to the *Human only* baseline condition—the GenAI conditions combined causally affect evaluators' third-party assessments and writers' self-assessments of the stories' creativity, in terms of two commonly used dimensions, novelty and usefulness (*12,18*).

We find that evaluators assess that stories composed by writers in the two GenAI conditions are more creative, in terms of both novelty and usefulness (Figure 2A; Figure S1 shows violin plot of raw data). We find that the provision of AI ideas improves the story's novelty by 6.7% ($b$=0.259, $p$=0.001, see Table S3; compared to the *Human only* mean of 3.85) and its usefulness by 6.4% ($b$=0.319, $p$<0.001; compared to the *Human only* mean of 5.02). Results for each of the constituent components of the novelty index (whether the story is novel, original, and rare) and usefulness index (whether the story is appropriate, feasible, and publishable) are consistent with the aggregate results for the indices (see Table S3).

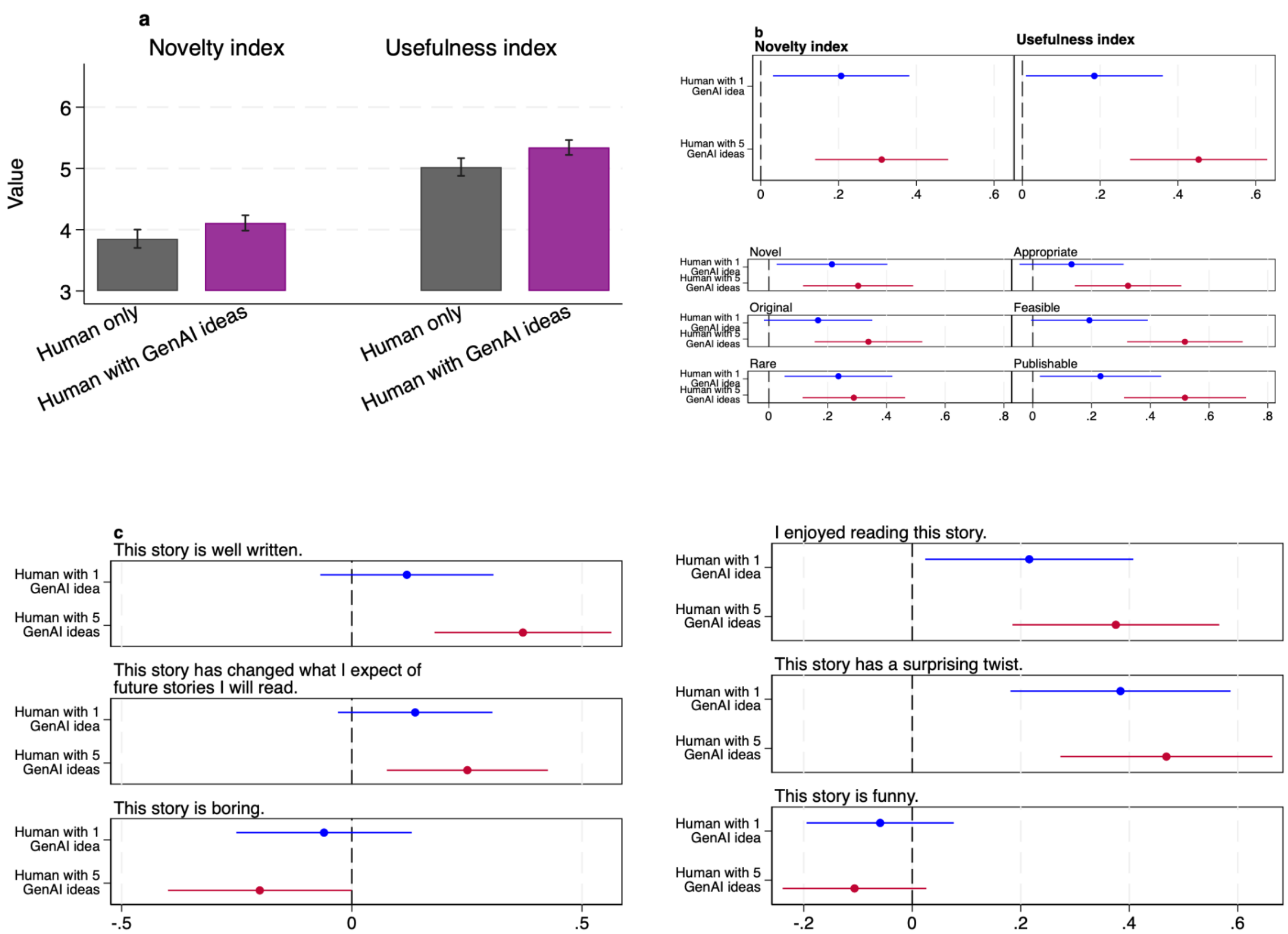

**Figure 2.** Evaluation of creativity and emotional characteristics by third party evaluators. **a**, Compares novelty and usefulness indices of participants in *Human only* condition to participants who had access to (1 or 5) GenAI ideas combined. **b**, Constituent components of each index. **c–d**, Compares novelty and usefulness indices (and constituent components) (**c**) and emotional characteristics (**d**) of *Human only* condition (dashed vertical line) to *Humans with 1 GenAI idea* and *Humans with 5 GenAI ideas* conditions.

**Variation by GenAI conditions**. To better understand how greater availability of GenAI ideas affects the enhancement in creativity, we follow our pre-registration to estimate the causal impact of the two GenAI conditions separately. Writers in the *Human with 1 GenAI idea* condition are given the choice to request a single GenAI story idea, while writers in the *Human with 5 GenAI ideas* condition are given the option to access up to five GenAI story ideas.

Across the two GenAI conditions, 88.4% participants choose to call upon GenAI at least once to provide an initial story idea. Of the 100 writers in the *Human with 1 GenAI idea* condition, 82 opted to generate one, while 93 out of 98 writers in the *Human with 5 GenAI Ideas* condition did so. When given the option to call upon GenAI more than once in the *Human with 5 GenAI ideas* condition, participants did so on average 2.55 times, with 24.5% requesting the maximum of five GenAI ideas.

We find that, while having access to one GenAI idea leads to somewhat greater creativity, the most gains (and significant differences in our pre-registered indices) come from writers who have access to five GenAI ideas (Figure 2B; Figure S2 shows violin plot of raw data). With respect to novelty, writers in the *Human with 1 GenAI idea* condition experienced an increase of 5.4% ($b$=0.207, $p$=0.021, see Table S4) over writers without GenAI access; whereas writers in the *Human with 5 GenAI ideas* condition showed an increase in novelty of 8.1% ($b$=0.311, $p$<0.001) over writers without GenAI access.

The results of story usefulness are even more striking. The usefulness of stories from writers with access to one GenAI idea is 3.7% ($b$=0.185, p=0.039) higher than that of writers with no GenAI access. Having access to up to five AI ideas increased usefulness by 9.0% ($b$=0.453, $p$<0.001) over those with no GenAI access and 5.1% ($p$=0.0012, compared to the *Human with 1 GenAI idea* mean of 5.21) over those with access to one GenAI idea. The overall results suggest that having access to more AI ideas lead to more creative storytelling. The novelty and creativity index results are qualitatively unchanged when we include evaluator fixed effects, story order fixed effects, story topic fixed effects, and an indicator variable that equals one if the writer accessed at least one GenAI idea (see Table S5).

In contrast, writers self-assessing their own stories showed no significant differences in the novelty and usefulness between authors who were offered GenAI ideas and those who were not (see Tables S6).

**Exploratory analyses: emotional characteristics**. Next, we turn to measures that gauge the evaluators' emotional responses to the stories, based on categories of general reader interest, including how well written, enjoyable, funny, and boring the stories are and the extent to which the story has a plot twist. We also ask whether the story changed the reader's expectations about future stories (based on literature theorist Robert Jauss' conception of more novel literature changing the reader's "horizon of expectations" in the future (*19*)).

As illustrated in Figure 2C, we find that stories written by writers with access to GenAI ideas are more enjoyable (*Human with 1 GenAI idea*: *b*=0.216, *p*=0.028; *Human with 5 GenAI ideas*: *b*=0.375, *p*<0.001, see Table S7) and are more likely to have plot twists (*Human with 1 GenAI idea*: *b*=0.384, *p*<0.001; *Human with 5 GenAI ideas*: *b*=0.468, *p*<0.001). Relative to *Human only* stories, when the writer has access to up to five GenAI ideas, the stories are considered to be better written (*b*=0.372, *p*<0.001), have more of an effect on the evaluator's expectations of future stories (*b*=0.251, *p*=0.005), and be less boring (*b*= –0.200, *p*=0.049). Stories in the *Human with 5 GenAI ideas* are, however, not evaluated as more funny than *Human only*; if anything the coefficient is negative but not significant (b= –0.106, *p*=0.115).

Again, writers' self-assessment of their own stories showed no significant differences in the story characteristics across conditions (see Tables S8).

**Heterogeneity by inherent creativity**. Since our human writers were not specifically selected for their creative predispositions or work in creative industries, we are able to take advantage of natural variation in the underlying creativity of writers in our sample. To do so, we had writers complete a Divergent Association Task (DAT) prior to writing their stories[17]. The task entails providing ten words that are as different from each other as possible. The DAT score is the cosine distance of the underlying word embeddings (scaled to 100) and captures the individual's inherent creativity. In our sample, the DAT score had a mean of 77.24 and a standard deviation of 6.48. The computation of DAT requires seven of ten submitted terms to be valid (i.e., single words that appear in the dictionary). Two writers failed to properly submit seven valid words, so the DAT score was successfully computed for 291 of 293 writers.

First, we look at whether different writers engage with GenAI more than others: we do not find differences between more creative writers and less creative writers in how frequently they access GenAI ideas in the two GenAI conditions (see Table S9). Among both more and less creative writers in the *Human with 5 GenAI ideas* condition, all five ideas were requested 24.5% of the time. In short, we do not observe any differences in how GenAI was accessed based on the inherent creativity of the writer.

Next, we interact the continuous DAT score with our conditions (see Tables S10 and S11 for results on all outcome variables). Figure 3 presents graphs that show the differential effect of GenAI ideas on select variables, based on the inherent creativity of the writer (see Figure S3 for graphs of remaining outcome variables). Among the most inherently creative writers (i.e., high-DAT writers), there is little effect of having access to GenAI ideas on the creativity of their stories. Across all conditions, high-DAT writers' stories are evaluated relatively highly, both in terms of novelty and usefulness; and providing them with access to GenAI does not affect their high evaluations. We observe a similar result among high-DAT writers for how well the story was written, how enjoyable and, conversely, how boring it is: having access to GenAI does not affect high-DAT writer's already good performance on these outcomes.

In contrast, access to GenAI ideas substantially improves the creativity and select emotional characteristics of stories written by inherently less creative writers (i.e., low-DAT writers). Among low-DAT writers, having access to 1 GenAI idea improves a story's novelty by 6.3% and having access to 5 GenAI ideas yields improvements of 10.7%. Similarly, writers with access to 1 and 5 GenAI ideas produce stories that are evaluated more highly on usefulness by 5.5% and 11.5%, respectively. Similar improvements exist for certain story characteristics. For low-DAT writers in the *Human with 5 GenAI ideas* condition, assessments of how well the story was written increase by up to 26.6%, enjoyment of the story increase by up to 22.6%, and how boring the story is decreases by up to 15.2%. These improvements in the creativity of low-DAT writers' stories put them on par with high-DAT writers. In short, the *Human with 5 AI Ideas* condition effectively equalizes the creativity scores across less and more creative writers.

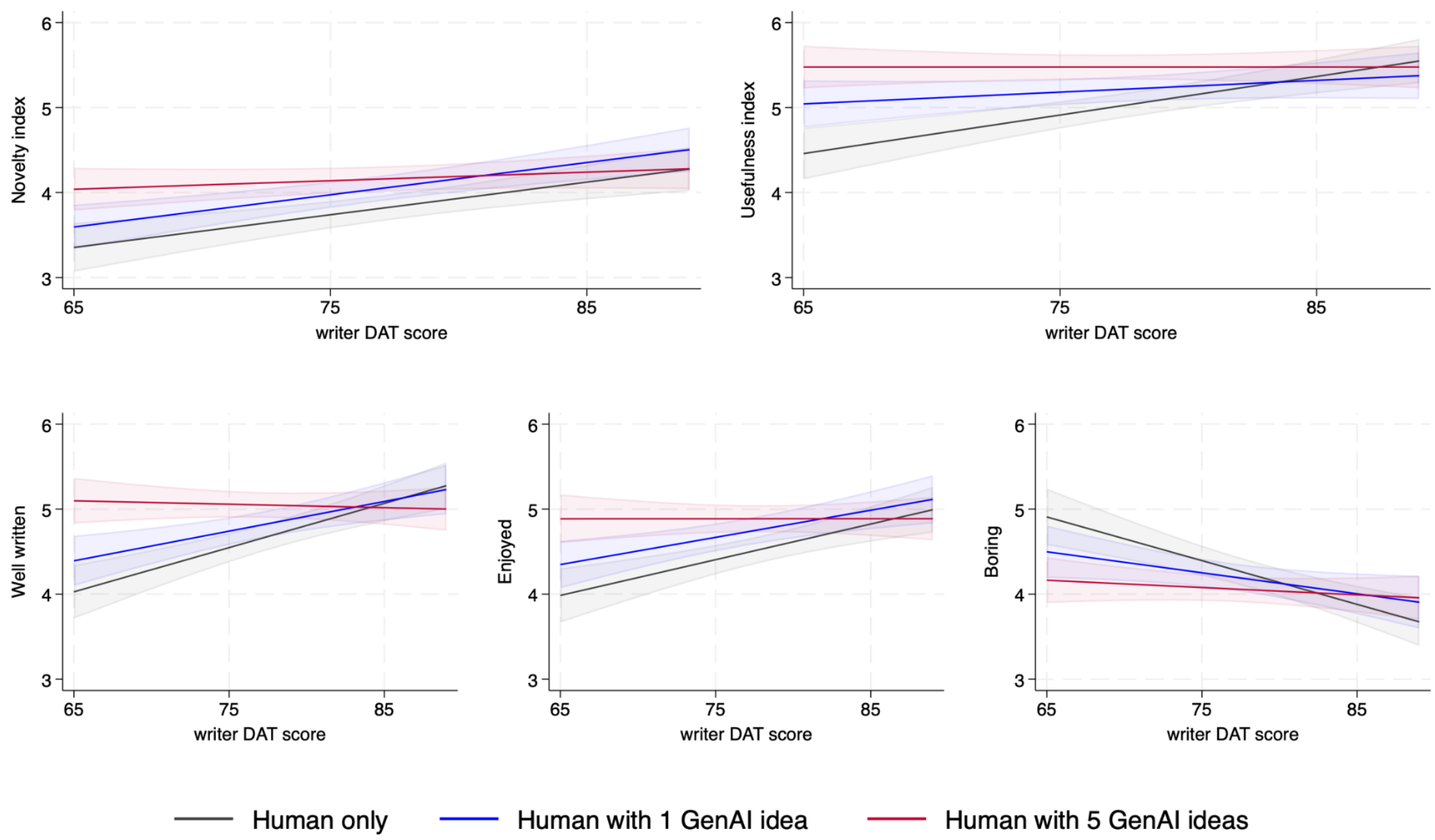

**Figure 3.** Marginal effect of writer's inherent creativity (as measured by DAT score) on the creativity indices and on select emotional characteristics by condition.

**Ownership and profit-sharing**. After asking evaluators to review each story for creativity and their emotional reactions, we ask evaluators to estimate to what extent each of the stories they just evaluated may have been influenced by GenAI (on a 100-point scale). We find that evaluators are able to assess whether stories received AI assistance (*Human with 1 GenAI idea*: $b$=6.21, $p$<0.001; *Human with 5 GenAI ideas*: $b$=4.96, $p$<0.001; see Table S12), though we do not find that they are more or less likely to attribute stories from writers with access to 5 GenAI ideas as being affected by GenAI, compared to those with access to 1 GenAI idea ($p$=0.3053).

Next, we disclose to evaluators whether GenAI ideas were made available to the writer of each story[13] and, if so, the GenAI ideas are shown alongside the final story. We ask evaluators how much ownership should be attributed to the writers for their final story, an index we compute based on their answers to questions about the extent to which a story reflects the writer's own ideas and their claim to ownership (Figure 4). Figure 4A shows that, for stories that were produced by writers who had access to GenAI, evaluators attribute substantially less ownership to the writer. Using the ownership index, evaluators ascribed 25.4% less ownership to authors who had access to one GenAI idea, relative to writers in the *Human only* condition ($b$ = –1.96, $p$<0.001, see Table S13; compared to the *Human only* mean of 7.74). The ownership discount is even higher for writers who had access to up to five GenAI ideas, at 31.0% ($b$ = –2.40, $p$<0.001).

Following the questions on ownership, we elicit beliefs from evaluators about how hypothetical profits from selling the short story should be shared. We ask this question only for stories in the *Human with 1 AI idea* and *Human with 5 AI ideas* conditions and only if the writer requested at least one GenAI idea. We ask evaluators to indicate what percent of the story's profits should belong to the writer of the story versus the creator of the GenAI tool. We find that evaluators only impose a marginally significant penalty of 2.3% to writers who had access to 5 GenAI ideas ($b$= –2.30, $p$=0.072, see Table S14), relative to having access to 1 GenAI idea. Furthermore, this weak relationship is no longer statistically significant when we include as a control variable the extent to which the evaluator ascribes ownership of the story to the writer in both conditions. A one standard deviation increase in the ownership index results in an additional 16.2% of profits allocated to the writers ($b$=7.70, $p$<0.001).

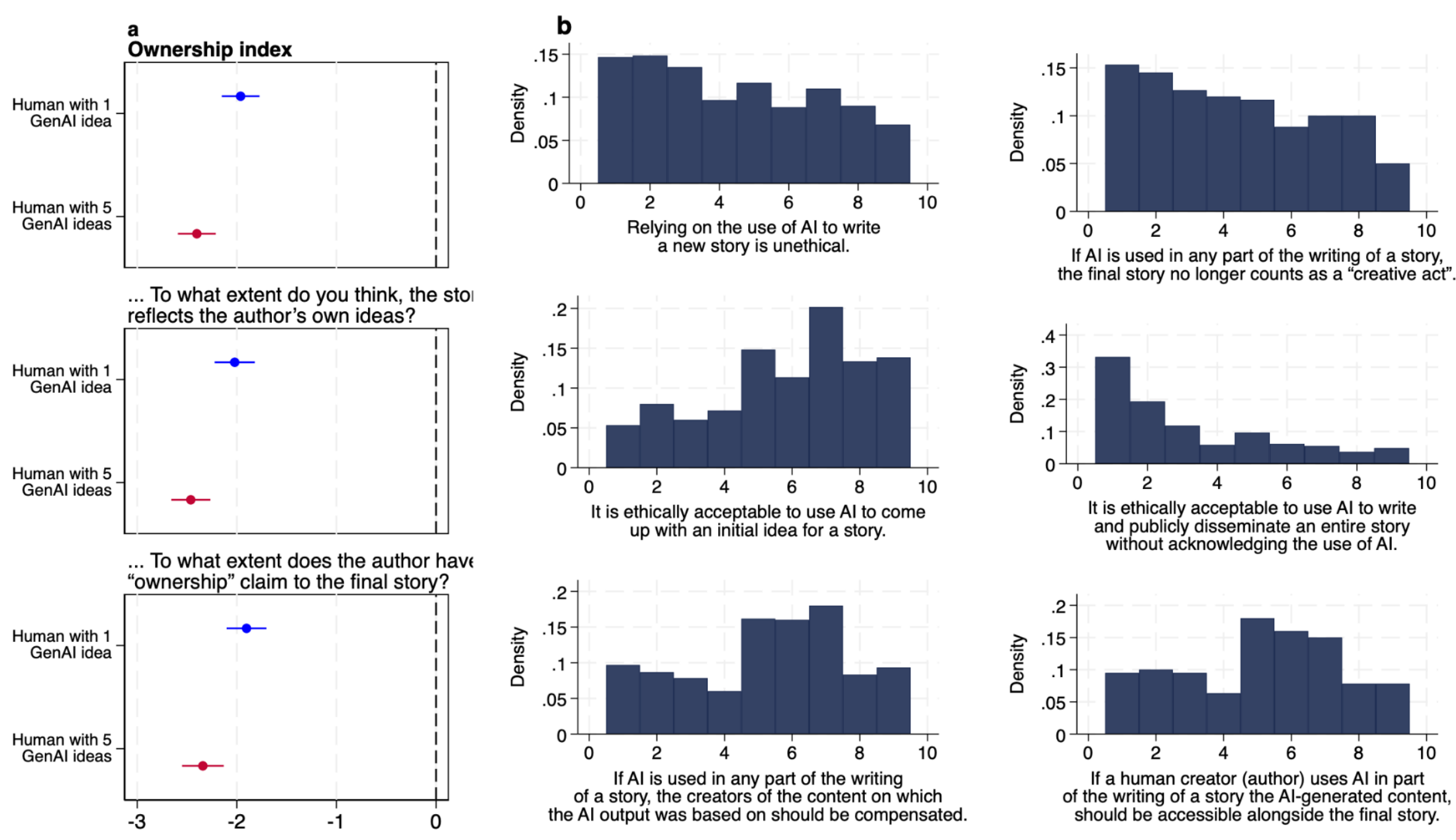

**Figure 4.** Evaluator assessment of ownership and attitudes toward genAI. **a**, Compares ownership index (and constituent components) of *Human only* (reference category) to humans with access to 1 and 5 GenAI ideas. **b**, summary of evaluator survey responses on attitudes toward genAI in creativity output.

**GenAI and ethics in the creative process.** We ask evaluators to indicate to what extent, and how, GenAI should be used to inspire stories in the future. We are interested in understanding the extent to which participants believe using GenAI is ethical and should be credited in the creative process. The responses to six exploratory questions are summarized in Figure 4B.

Evaluators in our sample tended to disagree with the ideas that the use of GenAI in story writing is unethical (52.7% scored 1 to 4 versus 35.7% scored 6 to 9; we focus on choices other than 5, since 5 is the scale midpoint, which might represent a neutral—i.e., indecisive—stance; see Tables S15 and S16) and that the story ceases to be a "creative act" if AI is used in any part of the story writing process (54.5% versus 33.8%).

However, according to evaluators, there were limits in the acceptability of the use of AI. While evaluators tended to agree that using AI for an initial idea was acceptable (26.5% versus 58.7%), they overwhelmingly tended to disagree with the idea that AI could be used for a story without acknowledgement of its use (70.2% versus 20.2%). There was also consensus that the content creators on which the AI output was based ought to be compensated (32.2% versus 51.7%) and, to a lesser extent, that the AI-generated content should be disclosed alongside the final story (35.3% versus 46.7%).

**Similarity of stories**. Thus far, we have focused on the subjective evaluation of third-party readers; now we turn to a more objective measure of the stories' content, in order to understand how GenAI affects the final stories produced. Using embeddings (*20*) obtained from OpenAI's embeddings application programming interface (API), we are able to compute the cosine similarity of the stories to the generated AI ideas as well as similarity to all other stories within condition (Figure 5). We multiply the cosine similarity score by 100 to arrive at a measure that ranges from zero to 100.

For stories in the *Human only* condition or in one of the GenAI idea conditions where the writer chose not to generate an idea, we randomly assigned an GenAI idea from the pool of ideas (that were created for other writers) within the same story topic. For writers in the two GenAI idea conditions who used the GenAI idea idea, we selected the first available idea. Then we tested how similar the stories were to the GenAI ideas. Relative to *Human only*, writers in the *Human with 1 GenAI idea* and *Human with 5 GenAI ideas* conditions wrote stories that were 5.2% ($b$=4.29, $p$<0.001, see Table S17; compared to a *Human only* mean of 82.85) and 5.0% ($b$=4.11, $p$<0.001) more similar to the GenAI ideas, respectively. In short, writers in the two GenAI conditions are anchored to some extent on the GenAI idea presented to them. To put these values in context, consider that in the *Human only* condition, similarity scores span a range of 17.67 points; therefore, a 4.29 increase in similarity from having access to one GenAI idea represents 24.3% of the total range.

We also look at the similarity of any one story to the "mass" of all stories within the same condition by computing the cosine similarity of the embedding of the focal story with the average embedding of all other stories in the same condition. Our results show that having access to GenAI ideas makes a story more similar to the average of other stories within the same condition (*Human with 1 GenAI idea* $b$=0.871, $p$<0.001; *Human with 5 GenAI ideas* $b$=0.718, $p$=0.003).

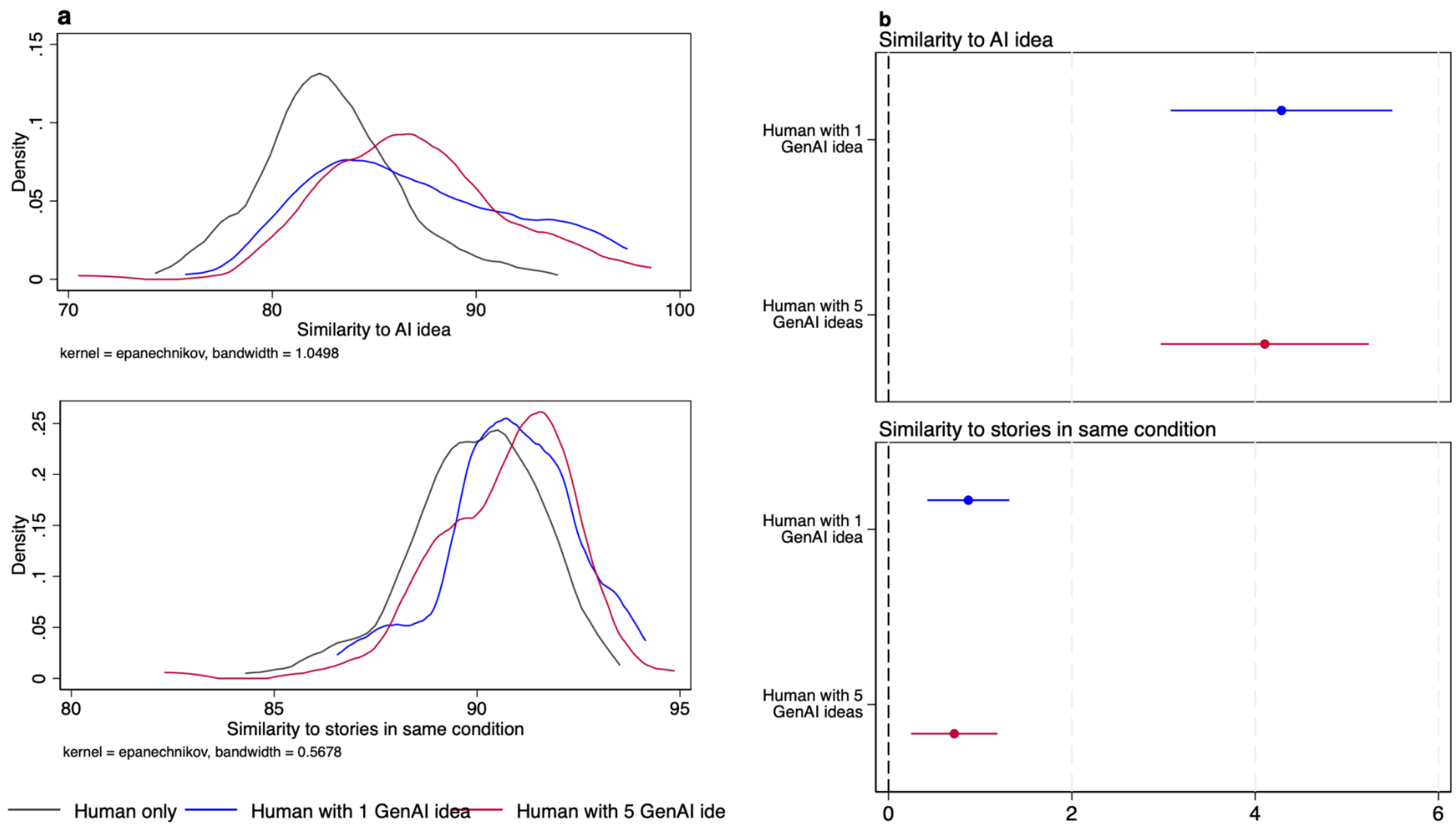

**Figure 5.** Comparison of similarity of writer stories to GenAI ideas and others stories. **a**, Kernel density plots comparing story similarity to ideas produced by GenAI and all other stories in the same condition for each condition. **b**, Compares story outcomes of *Human only* (reference category) to humans with access to 1 and 5 GenAI ideas.

## Discussion

GenAI has the potential to dramatically impact most aspects of the economy and society at large (*21,22*). Previous empirical work has focused on its effects on productivity, routine tasks, sales, resume writing, and joint collaboration between humans and AI, including for scientific and medical tasks (*3*–*6,23*–*25*), all of which contribute to our understanding of the potentially transformative impact of GenAI. In this paper, we extend this work by studying a question fundamental to all human behavior, which is of both economic and purely expressive value: how does generative AI affect human creativity?

We offer an answer to this question by experimentally studying the causal effect of having access to GenAI on writing short stories in an online experiment. We find that having access to GenAI causally increases the average novelty and usefulness—two frequently studied dimensions of creativity—relative to human writers on their own. This is driven, in particular, by our experimental condition that enables writers to request multiple GenAI ideas—up to five in our study—each presenting a different starting point, leading to a "tree" branching off to potential storylines (*3*).

Our results provide insight into how GenAI enhances creativity. Having access to GenAI "professionalizes" the stories beyond what writers might have otherwise accomplished alone. The overall effect is a more novel, more useful story that is well written and enjoyable. However, the gains from writing more creative stories benefit some more than others: less creative writers experience greater uplifts for their stories, seeing increases of 10% to 11% for creativity and of 22% to 26% for how enjoyable and well written the story is.

We note two additional observations about these results. First, having access to GenAI effectively equalizes the evaluations of stories, removing any disadvantage or advantage based on the writers' inherent creativity (*26*). That GenAI particularly benefited less able writers is paralleled in recent studies focusing on other domains in which GenAI has been shown to help less productive workers (*4,5*). Second, one might ask whether the GenAI ideas can push the upper bound of creativity of produced stories, beyond what particularly creative humans are capable of on their own. We do not find evidence of this possibility in this study; however, to test this hypothesis more thoroughly, we encourage future research to allow creatives to engage with GenAI more deeply and interactively, and to extend this work by leveraging more specialized GenAI tools for this purpose.

Our choice of the experimental design offers a fairly stringent test to measure the causal impact of GenAI on creativity. We designed our study such that endogenous decisions by the writer are minimized, but not fully eliminated. We do not allow writers to customize the call to the GenAI engine, nor do we allow for repeated interactions between writers and GenAI, both of which may increase the effectiveness and magnitude of the impact of GenAI on creativity. If that is the case, our estimates are likely a lower bound of the potential that GenAI could offer to writers when they are given full control over the AI engine, or when real-time interactions are enabled that help writers with ideation and enhancement further (*27*). Indeed, previous research suggests that conversations with others can improve ideation (*28*)—a possibility that GenAI could facilitate and enable in new and faster ways than ever before. That a tightly-controlled prompt requesting a GenAI idea shows significant effects on creativity in our study provides a promising starting point for future researchers to delve deeper into customization and personalization of GenAI for different writers (*8*).

We do, however, allow writers to opt into receiving GenAI ideas, rather than assign GenAI ideas to everyone in the GenAI conditions. We do this to ensure that writers are invested in, and receptive to, what GenAI produces. Furthermore, we anticipated that—if offered—the vast majority of participants would take advantage of the option to at least see the GenAI idea, thus minimizing the risk of self-selection affecting our causal estimates. The empirical evidence shows that nearly 9 out of 10 people in the GenAI conditions choose to receive at least one GenAI when offered, bolstering our confidence that our results—based on our conservative intention-to-treat analysis that studies the effect of condition regardless of whether writers did or did not choose to request GenAI ideas—allow for a causal interpretation.

Much has been written about the potential replacement of AI on human labor (e.g. automation) (*29,30*) or a “horse race” between human and AI-generated ideas (*31*). We focus on the potential complementarities of AI on human creative production. We do so among a sample of relatively “typical” study participants often used in academic studies (which comes with limitations on population representativeness) (*32*)—that is, we do not study professional writers or unusually creative individuals. These individuals remain an important but understudied population segment, for which the effects of GenAI could be transformative in other ways, potentially offering efficiency gains or improved speed of execution (*6*). That said, our results suggest that GenAI may have the largest impact on individuals who are less creative.

Despite the enhancement effect that GenAI had on average creativity, we also believe our study offers a cautionary note if GenAI were adopted more widely for creative tasks. In general equilibrium, an interesting question is whether the stories enhanced and inspired by AI will be able to create sufficient variation in the outputs they lead to. Specifically, if the publishing (and self-publishing) industry were to embrace more GenAI-inspired stories, our findings suggest that the produced stories would become less unique in aggregate and more similar to each other—an effect that may be heightened over time, as many GenAI models use the existing corpus of text and published literature to train and refine their algorithms and predictions, thus further reducing diversity. Initial evidence suggests that GenAI models trained with GenAI content become unstable (*33*).

GenAI is a rapidly evolving technology with its full potential yet to be explored. While our study used the most recent version of a widely used GenAI tool—OpenAI’s GPT-4—this may soon become obsolete. However, rather than limiting our study or future studies, we believe the fast progress of GenAI development offers exciting opportunities for researchers interested in creativity, art, literature, and innovation. If GenAI leads to enhancements of human creativity in a conservatively designed experimental study today, the creative possibilities for tomorrow may extend beyond our current, collective imagination.

## Methods

**Writer study and experimental conditions**. For the Writer Study, we recruited 500 participants to participate in the experiment from the Prolific platform. Using the platform’s filtering options, we included participants who were Prolific participants who indicated that they are based in the United Kingdom with an approval rating of at least 95% from between 100 and 1,000,000 prior submissions. Writers are not selected based on prior writing skills or their creativity. Of the 500 participants who began the study, 169 exited the study prior to giving consent, 22 were dropped for not giving consent, and 13 dropped out prior to completing the study. Three participants in the *Human only* condition admitted to using GenAI during their story writing exercise and—as per our pre-registration—they were therefore dropped from the analysis, resulting in a total number of writers and stories of 293.

We first ask each participant to complete the divergent association task (*17*), a trait measure of creativity. Each participant is then provided with instructions to complete a story writing task. Participants are randomized into writing about one of the following three topics: an adventure on the open seas, an adventure in the jungle, and an adventure on a different planet. Participants (using the "open seas" writing topic as an example) receive the following instructions: "We would like you to write a story about an adventure on the open seas. You can write about anything you like. The story must be exactly eight sentences long and it needs to be written in English and appropriate for a teenage and young adult audience (approximately 15 to 24 years of age)."

Participants are randomized into one of three experimental conditions: *Human only*, *Human with 1 GenAI idea*, and *Human with 5 GenAI ideas*. In the *Human only* condition, the participant is provided with a text box in which she can provide her response. Automatic checks are conducted to ensure the story meets the length requirements of eight sentences before the participant can continue. In the *Human with 1 GenAI idea* and the *Human with 5 GenAI ideas* conditions, the participant has the option to receive a three-sentence idea for a story from a GenAI tool. When a participant clicks on "Generate Story Idea…", we pass the following prompt to OpenAI's GPT API (again, using the open seas topic as an example): "Write a three-sentence summary of a story about an adventure on the open seas." The response from the API is passed to the participant. At the time of the study, we used the API from OpenAI's latest model, GPT-4. Those in the *Human with 1 GenAI idea* condition can only receive one story idea, while those in the *Human with 5 GenAI ideas* condition can receive up to five story ideas, each of which is visible to the participant. Screenshots of the interface presented to writer participants in each of the three conditions are shown in SI Section 5. Participants were not able to copy and paste the GenAI idea text.

We then asked the writers to evaluate their own stories. First, we asked them how much they agreed with six stylistic statements, including whether they enjoyed writing it, how well written it was, how boring it is, how funny it is, to what extent there was a surprise twist, and whether it changed their expectations of future stories (questions were asked in a random order across participants). We then asked participants about their view of story profits they should receive (as a percentage) and whether the story reflected their own ideas, as well as the novelty and usefulness of the story (on a 9-point scale). We also ask the *Human only* condition whether they used AI to help them complete the task. (As described above, if writers in the *Human only* condition answer "yes" to this question, they are not included in our main analysis, as per our pre-registration.)

**Evaluator study.** For the Evaluator Study, the 293 total stories were then evaluated by a separate set of evaluators on Prolific who did not take part in the writer study. Using the platform's filtering options, we included participants who were Prolific participants who indicated that they are based in the United Kingdom with an approval rating of at least 95% from between 100 and 1,000,000 prior submissions and had not previously participated in the Writer Study. Participants were not selected on the basis of prior experience in the publishing industry, but represent "regular" readers. Each evaluator was shown 6 stories (2 stories from each topic). The evaluations associated with the writers who did not complete the writer study and those in the *Human only* condition who acknowledged using AI to complete the story were dropped. Thus, there are a total of 3,519 evaluations of 293 stories made by 600 evaluators. Four evaluations remained for five evaluators, five evaluations remained for 71, and all six remained for 524 evaluators.

The order in which the authors reviewed stories was randomized across evaluators. Evaluators were presented with one story at a time and asked to provide their feedback on the stylistic characteristics, novelty, and usefulness of the story. We presented the evaluator the same stories a second time and asked for an assessment of whether the story was written by a human or AI (as a percentage). We then disclosed whether the writer was offered the opportunity to generate an AI idea and, if so, whether the writer made use of it. If the author did use AI, we provide the evaluator with the text of the idea. Following that disclosure, we asked about the extent to which the story reflects the author's ideas and the extent to which the author has an ownership claim over the story. If the author used AI, we also ask the share of the profit the author should receive. After the evaluations, we ask participants to assess six statements about the use of AI in writing stories. Screenshots of the interface presented to evaluator participants are shown in SI Section 6.

**Outcome variables.** For our pre-registered indices, we follow Harvey and Berry's (2022) definition of creativity in terms of novelty and usefulness (*12*), which draws on a diverse range of interpretations of creativity in the literature. Unless otherwise noted, all outcome (dependent) variables are assessed on a 9-point scale from 1 (not at all) to 9 (extremely) to capture disagreement versus agreement with a statement or a question. The exact wording for each statement or question is shown in SI Sections 5 and 6.

*Creativity.* Our novelty index has three components (novel, original, and rare), with which we create an average value. The usefulness index also has three components (appropriate, feasible, and publishable), with which we also create an average value. Furthermore, we explore six additional outcome variables focused on how enjoyable, how well written, how boring, and how funny the story is, as well as whether the story has a surprising twist and whether it has changed what the reader expects of future stories.

*Characteristics, ownership and profits.* Next, evaluators indicate the extent to which they believe each story was based on inputs from a GenAI tool (on a scale from 0% to 100%). On the following pages, they learn if GenAI was available to writers and then state the extent to which the writer has ownership over the final story and it reflects the author's own ideas. These two questions are averaged to create an ownership index. In addition, if GenAI was used, evaluators are also asked to choose how to split hypothetical profits between the writer and the creator of the AI tool (on a scale from 0% to 100%).

*Ethics and use of AI.* Finally, evaluators also indicate across six statements their beliefs about ethical uses of AI in producing creative out, including how ethical the use of AI is, to what extent a story using AI would still count as a "creative act", content creators on which the AI idea is based should be compensated, AI should be credited and whether the AI input should be accessible alongside the final story.

*Similarity scores.* We compute measures of the writer's story to a GenAI idea as well as to all other stories from writers in the same condition. We do so by computing the cosine similarity of the embeddings and multiplying the value by 100 to arrive at a measure that ranges from zero to 100. Embeddings are obtained via a call to OpenAI's embeddings API. For GenAI ideas, we first randomly assign a GenAI story from the same condition amongst all GenAI ideas to all writers who did not have an idea (i.e., all writers in the *Human only* condition and writers in the GenAI idea conditions who opted not to request for any GenAI ideas). For writers who opted to receive multiple GenAI ideas, we selected the first available idea. First, we compute the cosine similarity of the embeddings of the story and the respective GenAI idea. Second, for the similarity measure to all other stories, we take the cosine similarity of the embedding of the focal story with the average embedding for all other stories in the same condition.

**Statistical analysis.** Unless otherwise noted, we run regressions using ordinary least squares (OLS) using robust standard errors for outcomes derived from the Writer Study (each writer produces one story) and robust standard errors clustered at the participant (i.e., evaluator) level for those derived from the Evaluator Study (each evaluator assesses six stories). The key independent variables are the conditions to which writers are exogenously assigned where *Human Only* is the baseline (reference category) and the *Human with 1 AI idea* and *Human with 5 AI ideas* conditions are dummy variables.

**Pre-registration and ethics approval.** The study was pre-registered at AsPredicted.org (ID 136723); a copy of the pre-registration is included in the SI Section 4. The study was approved by the ethics boards at UCL (ID UCLSOM-2023-002) and the University of Exeter (ID 1642263).

**Acknowledgements.** We are grateful to Scott Vincent for research assistance and programming support. We also thank participants at various paper presentations and panel sessions at the 2023 Academy of Management.

*Supplemental Information for:*

# Generative artificial intelligence enhances creativity but reduces the diversity of novel content

**Anil R. Doshi** and **Oliver P. Hauser**

**Table of contents**

# Section 1. Summary of questions for Writer and Evaluator studies

Participants are asked to indicate to what extent they agree with each statement or question on scale from 1 (not at all) to 9 (extremely).

| | **Question text** [blue = evaluator, red = writer] | **Part of an index?** | **Asked of writers?** | **Asked of evaluators?** |
|---|---|---|---|---|
| 1 | How novel do you think the / your story is? | Novelty index | ✓ | ✓ |
| 2 | How original do you think the / your story is? | Novelty index | ✓ | ✓ |
| 3 | How rare (i.e., unusual) do you think the / your story is? | Novelty index | ✓ | ✓ |
| 4 | How appropriate do you think the / your story is for the intended audience? | Usefulness index | ✓ | ✓ |
| 5 | How feasible do you think the / your story is to be developed into a complete book? | Usefulness index | ✓ | ✓ |
| 6 | How likely do you think it would be that the / your story is turned into a complete book if a publisher read it and hired a professional author to expand on the idea? | Usefulness index | ✓ | ✓ |
| 7 | I enjoyed reading / writing this story. | | ✓ | ✓ |
| 8 | This story is well written. | | ✓ | ✓ |
| 9 | This story is boring. | | ✓ | ✓ |
| 10 | This story has changed what I expect of future stories I will read. | | ✓ | ✓ |
| 11 | This story is funny. | | ✓ | ✓ |
| 12 | This story has a surprising twist. | | ✓ | ✓ |
| 13 | Please indicate to what extent this specific AI generated idea affected the story you submitted. | | ✓ | |
| 14 | Please indicate the extent (if any) to which you think this story was based on inputs from an AI tool (e.g. ChatGPT or similar generative AI tool). (0% to 100% scale) | | | ✓ |
| 15 | To what extent do you think the / your story reflects the author's / your own ideas? | Ownership index (Evaluators) | ✓ | ✓ |
| 16 | To what extent does the author have an "ownership" claim to the final story? | Ownership index (Evaluators) | | ✓ |
| 17 | If this story were published and sold tomorrow, how much of the story's profit do you believe should belong to (you / the author) versus the creators of the generative AI tool that may have provided the starting point for the story? (0% to 100% scale) | | | ✓ |
| 18 | Relying on the use of AI to write a new story is unethical. | | | ✓ |
| 19 | If AI is used in any part of the writing of a story, the final story no longer counts as a "creative act". | | | ✓ |
| 20 | It is ethically acceptable to use AI to come up with an initial idea for a story. | | | ✓ |
| 21 | It is ethically acceptable to use AI to write an entire story without acknowledging the use of AI. | | | ✓ |
| 22 | If AI is used in any part of the writing of a story, the creators of the content on which the AI output was based on should be compensated. | | | ✓ |
| 23 | If a human creator (author) uses AI in part of the writing of a story, the AI-generated content should be accessible alongside the final story. | | | ✓ |

# Section 2. Supporting Tables and Figures

## Supplementary Table 1. Comparison of Means for writers

| | Human | 1AI idea | 5AI ideas | Human / 1AI idea | Human / 5AI ideas | 1AI idea / 5AI ideas |
|---|---|---|---|---|---|---|
| | mean | mean | mean | p | p | p |
| writer DAT score | 77.617 | 76.868 | 77.254 | (0.415) | (0.698) | (0.683) |
| writer creative | 5.505 | 5.740 | 5.673 | (0.473) | (0.601) | (0.832) |
| writer creative job | 4.747 | 4.160 | 4.612 | (0.109) | (0.707) | (0.203) |
| writer tech comfort | 7.147 | 6.870 | 6.878 | (0.290) | (0.252) | (0.978) |
| writer AI engagement | 4.232 | 4.870 | 4.796 | (0.061) | (0.093) | (0.823) |
| writer used ChatGPT | 0.484 | 0.630 | 0.561 | (0.041) | (0.287) | (0.327) |
| writer used text AI tools | 0.474 | 0.590 | 0.571 | (0.105) | (0.176) | (0.792) |
| writer used image AI tools | 0.274 | 0.300 | 0.265 | (0.686) | (0.896) | (0.590) |
| writer used audio AI tools | 0.053 | 0.110 | 0.061 | (0.143) | (0.798) | (0.222) |
| writer used music AI tools | 0.053 | 0.060 | 0.041 | (0.824) | (0.699) | (0.539) |
| writer used video AI tools | 0.042 | 0.020 | 0.020 | (0.379) | (0.391) | (0.984) |
| writer gender female | 0.379 | 0.440 | 0.408 | (0.389) | (0.680) | (0.652) |
| writer income > £50,000 | 0.158 | 0.080 | 0.133 | (0.095) | (0.621) | (0.232) |
| writer education undergraduate or more | 0.168 | 0.170 | 0.204 | (0.977) | (0.527) | (0.541) |
| writer employed part- or full-time | 0.779 | 0.750 | 0.724 | (0.636) | (0.384) | (0.685) |
| writer age | 38.526 | 41.050 | 39.041 | (0.170) | (0.785) | (0.282) |
| Observations | 95 | 100 | 98 | 195 | 193 | 198 |

## Supplementary Table 2. Comparison of means for evaluators (selecting on condition of first story evaluated)

| | Human | 1AI idea | 5AI ideas | Human / 1AI idea | Human / 5AI ideas | 1AI idea / 5AI ideas |
|---|---|---|---|---|---|---|
| | mean | mean | mean | p | p | p |
| evaluator creative | 5.648 | 5.522 | 5.580 | (0.562) | (0.749) | (0.793) |
| evaluator creative job | 4.528 | 4.507 | 4.395 | (0.936) | (0.599) | (0.654) |
| evaluator tech comfort | 7.171 | 7.030 | 6.935 | (0.370) | (0.135) | (0.556) |
| evaluator AI engagement | 4.598 | 4.368 | 4.420 | (0.351) | (0.468) | (0.830) |
| evaluator used ChatGPT | 0.633 | 0.602 | 0.600 | (0.522) | (0.497) | (0.968) |
| evaluator used text AI tools | 0.643 | 0.562 | 0.625 | (0.098) | (0.707) | (0.201) |
| evaluator used image AI tools | 0.347 | 0.333 | 0.285 | (0.778) | (0.186) | (0.296) |
| evaluator used audio AI tools | 0.075 | 0.065 | 0.070 | (0.676) | (0.837) | (0.832) |
| evaluator used music AI tools | 0.055 | 0.050 | 0.075 | (0.805) | (0.426) | (0.297) |
| evaluator used video AI tools | 0.075 | 0.040 | 0.045 | (0.128) | (0.203) | (0.797) |
| evaluator gender female | 0.467 | 0.468 | 0.480 | (0.995) | (0.801) | (0.805) |
| evaluator income > £50,000 | 0.141 | 0.174 | 0.165 | (0.360) | (0.501) | (0.808) |
| evaluator education undergraduate or more | 0.211 | 0.214 | 0.170 | (0.944) | (0.298) | (0.265) |
| evaluator employed part- or full-time | 0.724 | 0.811 | 0.750 | (0.039) | (0.551) | (0.141) |
| evaluator age | 40.206 | 39.697 | 39.065 | (0.692) | (0.380) | (0.623) |
| Observations | 199 | 201 | 200 | 400 | 399 | 401 |

## Supplementary Table 3. Evaluator assessment of creativity (combined AI idea conditions)

| | (1) | (2) | (3) | (4) | (5) | (6) | (7) | (8) |
|---|---|---|---|---|---|---|---|---|
| | Novelty index | Novel | Original | Rare | Usefulness index | Appropriate | Feasible | Publishable |
| Human with GenAI idea(s) | 0.259** | 0.259** | 0.253** | 0.263** | 0.319*** | 0.228** | 0.355*** | 0.374*** |
| | (0.078) | (0.085) | (0.082) | (0.080) | (0.080) | (0.080) | (0.089) | (0.094) |
| Constant | 3.851*** | 4.023*** | 3.972*** | 3.559*** | 5.023*** | 5.708*** | 4.810*** | 4.551*** |
| | (0.076) | (0.083) | (0.081) | (0.078) | (0.073) | (0.078) | (0.082) | (0.086) |
| Observations | 3519 | 3519 | 3519 | 3519 | 3519 | 3519 | 3519 | 3519 |
| F-Stat | 10.9 | 9.26 | 9.47 | 10.7 | 16.0 | 8.04 | 16.0 | 16.0 |
| Adj R-squared | 0.0032 | 0.0028 | 0.0025 | 0.0031 | 0.0048 | 0.0022 | 0.0047 | 0.0050 |

Note: + $p < 0.10$, * $p < 0.05$, ** $p < 0.01$, *** $p < 0.001$.

## Supplementary Table 4. Evaluator assessment of creativity (separate AI idea conditions)

| | (1) | (2) | (3) | (4) | (5) | (6) | (7) | (8) |
|---|---|---|---|---|---|---|---|---|
| | Novelty index | Novel | Original | Rare | Usefulness index | Appropriate | Feasible | Publishable |
| Human with 1 GenAI idea | 0.207* | 0.215* | 0.168+ | 0.237* | 0.185* | 0.132 | 0.193+ | 0.230* |
| | (0.089) | (0.096) | (0.094) | (0.093) | (0.090) | (0.090) | (0.101) | (0.105) |
| Human with 5 GenAI ideas | 0.311*** | 0.304** | 0.339*** | 0.289** | 0.453*** | 0.324*** | 0.518*** | 0.518*** |
| | (0.087) | (0.095) | (0.093) | (0.089) | (0.090) | (0.092) | (0.100) | (0.106) |
| Constant | 3.851*** | 4.023*** | 3.972*** | 3.559*** | 5.023*** | 5.708*** | 4.810*** | 4.551*** |
| | (0.076) | (0.083) | (0.081) | (0.078) | (0.073) | (0.078) | (0.082) | (0.086) |
| Observations | 3519 | 3519 | 3519 | 3519 | 3519 | 3519 | 3519 | 3519 |
| F-Stat | 6.42 | 5.18 | 6.65 | 5.68 | 13.2 | 6.35 | 14.1 | 12.3 |
| Adj R-squared | 0.0033 | 0.0028 | 0.0032 | 0.0029 | 0.0073 | 0.0032 | 0.0075 | 0.0070 |

Note: + $p < 0.10$, * $p < 0.05$, ** $p < 0.01$, *** $p < 0.001$.

## Supplementary Table 5. Evaluator assessment of creativity (robustness checks)

| | (1) | (2) | (3) | (4) | (5) | (6) | (7) | (8) |
|---|---|---|---|---|---|---|---|---|
| | Novelty index | Novelty index | Novelty index | Novelty index | Usefulness index | Usefulness index | Usefulness index | Usefulness index |
| Human with 1 GenAI idea | 0.204** | 0.203** | 0.131+ | 0.246+ | 0.246** | 0.248** | 0.215* | 0.243+ |
| | (0.079) | (0.078) | (0.077) | (0.134) | (0.084) | (0.083) | (0.083) | (0.135) |
| Human with 5 GenAI ideas | 0.355*** | 0.354*** | 0.322*** | 0.455** | 0.538*** | 0.540*** | 0.536*** | 0.569*** |
| | (0.078) | (0.077) | (0.076) | (0.146) | (0.084) | (0.084) | (0.084) | (0.149) |
| Used AI | | | | -0.142 | | | | -0.035 |
| | | | | (0.128) | | | | (0.129) |
| Constant | 3.837*** | 3.590*** | 3.561*** | 3.561*** | 4.974*** | 5.145*** | 5.198*** | 5.198*** |
| | (0.047) | (0.076) | (0.084) | (0.084) | (0.050) | (0.083) | (0.095) | (0.095) |
| Story order fixed effects | No | Yes | Yes | Yes | No | Yes | Yes | Yes |
| Story topic fixed effects | No | No | Yes | Yes | No | No | Yes | Yes |
| Evaluator fixed effects | Yes | Yes | Yes | Yes | Yes | Yes | Yes | Yes |
| Observations | 3519 | 3519 | 3519 | 3519 | 3519 | 3519 | 3519 | 3519 |
| F-Stat | 10.4 | 6.08 | 11.5 | 10.5 | 21.3 | 7.02 | 6.43 | 5.79 |
| Adj R-squared | 0.0068 | 0.012 | 0.033 | 0.033 | 0.014 | 0.016 | 0.019 | 0.019 |

Note: + $p < 0.10$, * $p < 0.05$, ** $p < 0.01$, *** $p < 0.001$.

## Supplementary Table 6. Writer self-evaluation of creativity

| | (1) | (2) | (3) | (4) | (5) | (6) | (7) | (8) |
|---|---|---|---|---|---|---|---|---|
| | writer novel index | story novel | story original | story rare | writer useful index | story appropriate | story feasible | story publishable |
| Human with 1 GenAI idea | -0.025 | -0.077 | -0.053 | 0.054 | 0.078 | 0.237 | -0.271 | 0.267 |
| | (0.292) | (0.318) | (0.335) | (0.317) | (0.251) | (0.200) | (0.352) | (0.365) |
| Human with 5 GenAI ideas | -0.206 | -0.196 | -0.455 | 0.033 | 0.170 | -0.018 | 0.189 | 0.339 |
| | (0.295) | (0.316) | (0.331) | (0.311) | (0.251) | (0.201) | (0.351) | (0.370) |
| Constant | 4.505*** | 4.737*** | 4.863*** | 3.916*** | 5.779*** | 7.263*** | 5.821*** | 4.253*** |
| | (0.214) | (0.236) | (0.244) | (0.223) | (0.176) | (0.135) | (0.249) | (0.263) |
| Observations | 293 | 293 | 293 | 293 | 293 | 293 | 293 | 293 |
| F-Stat | 0.30 | 0.20 | 1.18 | 0.015 | 0.23 | 0.94 | 0.87 | 0.47 |
| Adj R-squared | -0.0048 | -0.0055 | 0.0010 | -0.0068 | -0.0053 | -0.00012 | -0.00082 | -0.0036 |

Note: + $p < 0.10$, * $p < 0.05$, ** $p < 0.01$, *** $p < 0.001$.

## Supplementary Table 7. Evaluator assessment of emotional characteristics

| | (1) | (2) | (3) | (4) | (5) | (6) |
|---|---|---|---|---|---|---|
| | Well written | Enjoyed | Funny | Future | Twist | Boring |
| Human with 1 GenAI idea | 0.120 | 0.216* | -0.059 | 0.138 | 0.384*** | -0.060 |
| | (0.096) | (0.098) | (0.069) | (0.085) | (0.103) | (0.097) |
| Human with 5 GenAI ideas | 0.372*** | 0.375*** | -0.106 | 0.251** | 0.468*** | -0.200* |
| | (0.098) | (0.097) | (0.067) | (0.089) | (0.100) | (0.102) |
| Constant | 4.677*** | 4.512*** | 2.085*** | 3.042*** | 3.414*** | 4.258*** |
| | (0.081) | (0.080) | (0.060) | (0.083) | (0.083) | (0.081) |
| Observations | 3519 | 3519 | 3519 | 3519 | 3519 | 3519 |
| F-Stat | 7.87 | 7.48 | 1.25 | 3.97 | 11.8 | 2.07 |
| Adj R-squared | 0.0040 | 0.0040 | 0.00019 | 0.0018 | 0.0065 | 0.00068 |

Note: + $p < 0.10$, * $p < 0.05$, ** $p < 0.01$, *** $p < 0.001$.

## Supplementary Table 8. Writer self-evaluation of emotional characteristics

| | (1) | (2) | (3) | (4) | (5) | (6) |
|---|---|---|---|---|---|---|
| | writer well written | writer enjoyed | writer funny | writer future | writer twist | writer boring |
| Human with 1 GenAI idea | 0.059 | -0.171 | -0.425 | -0.274 | 0.177 | 0.022 |
| | (0.281) | (0.257) | (0.289) | (0.286) | (0.376) | (0.300) |
| Human with 5 GenAI ideas | -0.146 | -0.153 | -0.401 | -0.161 | 0.272 | 0.276 |
| | (0.280) | (0.232) | (0.282) | (0.299) | (0.365) | (0.288) |
| Constant | 5.421*** | 7.011*** | 3.105*** | 3.274*** | 4.453*** | 3.968*** |
| | (0.197) | (0.183) | (0.210) | (0.206) | (0.268) | (0.208) |
| Observations | 293 | 293 | 293 | 293 | 293 | 293 |
| F-Stat | 0.28 | 0.28 | 1.36 | 0.46 | 0.28 | 0.57 |
| Adj R-squared | -0.0049 | -0.0048 | 0.0029 | -0.0039 | -0.0050 | -0.0031 |

Note: + $p < 0.10$, * $p < 0.05$, ** $p < 0.01$, *** $p < 0.001$.

# Supplementary Table 9. Writer creativity and accessing GenAI Ideas

| | (1) | (2) |
|---|---|---|
| | Used AI | Used AI |
| Model: | OLS | Logistic |
| Human with 5 GenAI ideas | 0.306 | 8.938 |
| | (0.483) | (8.178) |
| writer DAT score | -0.003 | -0.021 |
| | (0.005) | (0.034) |
| Human with 5 GenAI ideas # writer DAT score | -0.002 | -0.094 |
| | (0.006) | (0.100) |
| Constant | 1.048** | 3.134 |
| | (0.349) | (2.617) |
| Observations | 197 | 197 |
| F-Stat / Wald Chi-squared (logistic) | 4.00 | 6.58 |
| Adj R-squared / Pseudo R-squared (logistic) | 0.034 | 0.081 |

Note: + $p < 0.10$, * $p < 0.05$, ** $p < 0.01$, *** $p < 0.001$.

## Supplementary Table 10. Evaluator assessment of creativity (conditions interacted with writer's DAT score)

| | (1) | (2) | (3) | (4) | (5) | (6) | (7) | (8) |
|---|---|---|---|---|---|---|---|---|
| | Novelty index | Novel | Original | Rare | Usefulness index | Appropriate | Feasible | Publishable |
| Human with 1 GenAI idea | 0.273 | 0.673 | -0.766 | 0.911 | 2.633* | 2.643* | 2.457* | 2.798* |
| | (1.014) | (1.095) | (1.081) | (1.027) | (1.090) | (1.153) | (1.184) | (1.250) |
| Human with 5 GenAI ideas | 2.528** | 2.621* | 2.378* | 2.586* | 3.966*** | 3.721*** | 3.844*** | 4.331*** |
| | (0.969) | (1.039) | (1.045) | (1.010) | (1.024) | (1.082) | (1.138) | (1.197) |
| writer DAT score | 0.038*** | 0.040*** | 0.035*** | 0.040*** | 0.045*** | 0.044*** | 0.044*** | 0.048*** |
| | (0.010) | (0.010) | (0.010) | (0.010) | (0.010) | (0.011) | (0.012) | (0.012) |
| Human with 1 GenAI idea # writer DAT score | -0.001 | -0.006 | 0.012 | -0.008 | -0.032* | -0.032* | -0.029+ | -0.033* |
| | (0.013) | (0.014) | (0.014) | (0.013) | (0.014) | (0.015) | (0.015) | (0.016) |
| Human with 5 GenAI ideas # writer DAT score | -0.028* | -0.030* | -0.026+ | -0.029* | -0.045*** | -0.044** | -0.043** | -0.049** |
| | (0.012) | (0.013) | (0.013) | (0.013) | (0.013) | (0.014) | (0.015) | (0.015) |
| Constant | 0.857 | 0.898 | 1.244 | 0.428 | 1.511+ | 2.344** | 1.380 | 0.808 |
| | (0.753) | (0.815) | (0.813) | (0.764) | (0.806) | (0.857) | (0.907) | (0.915) |
| Observations | 3494 | 3494 | 3494 | 3494 | 3494 | 3494 | 3494 | 3494 |
| F-Stat | 9.18 | 7.46 | 9.43 | 8.58 | 8.72 | 5.26 | 8.41 | 8.20 |
| Adj R-squared | 0.012 | 0.010 | 0.012 | 0.011 | 0.013 | 0.0076 | 0.012 | 0.012 |

Note: + $p < 0.10$, * $p < 0.05$, ** $p < 0.01$, *** $p < 0.001$.

## Supplementary Table 11. Evaluator assessment of emotional characteristics (conditions interacted with DAT score)

| | (1) | (2) | (3) | (4) | (5) | (6) |
|---|---|---|---|---|---|---|
| | Well written | Enjoyed | Funny | Future | Twist | Boring |
| Human with 1 GenAI idea | 1.477 | 1.017 | -1.804* | 0.983 | 1.760 | -2.147+ |
| | (1.166) | (1.119) | (0.739) | (1.092) | (1.211) | (1.216) |
| Human with 5 GenAI ideas | 4.717*** | 3.629*** | 0.488 | 1.915+ | 2.461* | -3.531** |
| | (1.075) | (1.083) | (0.766) | (1.042) | (1.105) | (1.128) |
| writer DAT score | 0.052*** | 0.042*** | 0.003 | 0.032** | 0.057*** | -0.051*** |
| | (0.011) | (0.011) | (0.007) | (0.010) | (0.011) | (0.011) |
| Human with 1 GenAI idea # writer DAT score | -0.017 | -0.010 | 0.023* | -0.011 | -0.018 | 0.027+ |
| | (0.015) | (0.014) | (0.010) | (0.014) | (0.016) | (0.016) |
| Human with 5 GenAI ideas # writer DAT score | -0.056*** | -0.042** | -0.007 | -0.021 | -0.026+ | 0.043** |
| | (0.014) | (0.014) | (0.010) | (0.013) | (0.014) | (0.014) |
| Constant | 0.644 | 1.255 | 1.846** | 0.582 | -1.005 | 8.253*** |
| | (0.828) | (0.828) | (0.586) | (0.802) | (0.843) | (0.876) |
| Observations | 3494 | 3494 | 3494 | 3494 | 3494 | 3494 |
| F-Stat | 9.95 | 7.68 | 4.34 | 4.80 | 15.8 | 5.80 |
| Adj R-squared | 0.013 | 0.010 | 0.0029 | 0.0060 | 0.019 | 0.0076 |

Note: + $p < 0.10$, * $p < 0.05$, ** $p < 0.01$, *** $p < 0.001$.

## Supplementary Table 12. Evaluator assessment of AI assistance in story writing

| | (1) |
|---|---|
| | AI assistance |
| Human with 1 GenAI idea | 6.207*** |
| | (1.308) |
| Human with 5 GenAI ideas | 4.955*** |
| | (1.376) |
| Constant | 42.363*** |
| | (1.115) |
| Observations | 3519 |
| F-Stat | 11.8 |
| Adj R-squared | 0.0067 |

Note: + $p < 0.10$, * $p < 0.05$, ** $p < 0.01$, *** $p < 0.001$.

## Supplementary Table 13. Evaluator assessment of ownership

| | (1) | (2) | (3) |
|---|---|---|---|
| | Ownership index | authors ideas | ownership claim |
| Human with 1 GenAI idea | -1.962*** | -2.021*** | -1.902*** |
| | (0.097) | (0.103) | (0.102) |
| Human with 5 GenAI ideas | -2.401*** | -2.462*** | -2.341*** |
| | (0.097) | (0.100) | (0.107) |
| Constant | 7.736*** | 7.628*** | 7.843*** |
| | (0.075) | (0.077) | (0.078) |
| Observations | 3519 | 3519 | 3519 |
| F-Stat | 332.4 | 325.9 | 266.9 |
| Adj R-squared | 0.20 | 0.20 | 0.17 |

Note: + $p < 0.10$, * $p < 0.05$, ** $p < 0.01$, *** $p < 0.001$.

## Supplementary Table 14. Evaluator assessment of profit for writer (versus AI)

| | (1) | (2) | (3) | (4) |
|---|---|---|---|---|
| | profit share | profit share | profit share | profit share |
| Human with 5 GenAI ideas | -2.300+ | -0.913 | -0.971 | -1.276 |
| | (1.276) | (1.067) | (1.045) | (1.143) |
| Ownership index | | 7.702*** | | |
| | | (0.377) | | |
| ownership claim | | | 7.267*** | |
| | | | (0.349) | |
| authors ideas | | | | 5.775*** |
| | | | | (0.417) |
| Constant | 61.009*** | 19.468*** | 20.548*** | 30.868*** |
| | (1.233) | (2.115) | (1.957) | (2.509) |
| Observations | 2089 | 2089 | 2089 | 2089 |
| F-Stat | 3.25 | 213.8 | 219.8 | 98.8 |
| Adj R-squared | 0.0010 | 0.31 | 0.34 | 0.19 |

Note: + $p < 0.10$, * $p < 0.05$, ** $p < 0.01$, *** $p < 0.001$.

## Supplementary Table 15. Summary statistics of survey responses

| | Mean | S.D. | 25th pctile | 50th pctile | 75th pctile |
|---|---|---|---|---|---|
| Relying on the use of AI to write a new story is unethical. | 4.45 | 2.54 | 2.00 | 4.00 | 7.00 |
| If AI is used in any part of the writing of a story, the final story no longer counts as a "creative act". | 4.37 | 2.49 | 2.00 | 4.00 | 6.50 |
| It is ethically acceptable to use AI to come up with an initial idea for a story. | 5.83 | 2.36 | 4.00 | 6.00 | 8.00 |
| It is ethically acceptable to use AI to write an entire story without acknowledging the use of AI. | 3.27 | 2.45 | 1.00 | 2.00 | 5.00 |
| If AI is used in any part of the writing of a story, the creators of the content on which the AI output was based on should be compensated. | 5.28 | 2.40 | 3.00 | 6.00 | 7.00 |
| If a human creator (author) uses AI in part of the writing of a story, the AI-generated content should be accessible alongside the final story. | 5.08 | 2.36 | 3.00 | 5.00 | 7.00 |

Note: $n$ = 600.

## Supplementary Table 16. Heatmap of survey response counts (by question)

| | Response level | | | | | | | | |
|---|---|---|---|---|---|---|---|---|---|
| | 1 | 2 | 3 | 4 | 5 | 6 | 7 | 8 | 9 |
| Relying on the use of AI to write a new story is unethical. | 88 | 89 | 81 | 58 | 70 | 53 | 66 | 54 | 41 |
| If AI is used in any part of the writing of a story, the final story no longer counts as a "creative act". | 92 | 87 | 76 | 72 | 70 | 53 | 60 | 60 | 30 |
| It is ethically acceptable to use AI to come up with an initial idea for a story. | 32 | 48 | 36 | 43 | 89 | 68 | 121 | 80 | 83 |
| It is ethically acceptable to use AI to write an entire story without acknowledging the use of AI. | 199 | 116 | 71 | 35 | 58 | 37 | 33 | 22 | 29 |
| If AI is used in any part of the writing of a story, the creators of the content on which the AI output was based on should be compensated. | 58 | 52 | 47 | 36 | 97 | 96 | 108 | 50 | 56 |
| If a human creator (author) uses AI in part of the writing of a story, the AI-generated content should be accessible alongside the final story. | 57 | 60 | 57 | 38 | 108 | 96 | 90 | 47 | 47 |

# Supplementary Table 17. Writer story similarity

| | (1) | (2) |
|---|---|---|
| | story AI Idea similarity (incl simulated ideas) | story similarity to other stories in condition |
| Human with 1 GenAI idea | 4.288*** | 0.871*** |
| | (0.614) | (0.227) |
| Human with 5 GenAI ideas | 4.105*** | 0.718** |
| | (0.577) | (0.240) |
| Constant | 82.850*** | 89.961*** |
| | (0.343) | (0.161) |
| Observations | 293 | 293 |
| F-Stat | 37.3 | 8.23 |
| Adj R-squared | 0.16 | 0.044 |

Note: + $p < 0.10$, * $p < 0.05$, ** $p < 0.01$, *** $p < 0.001$.

Supplementary Figure 1. Violin plot of Human only and GenAI conditions (combined)

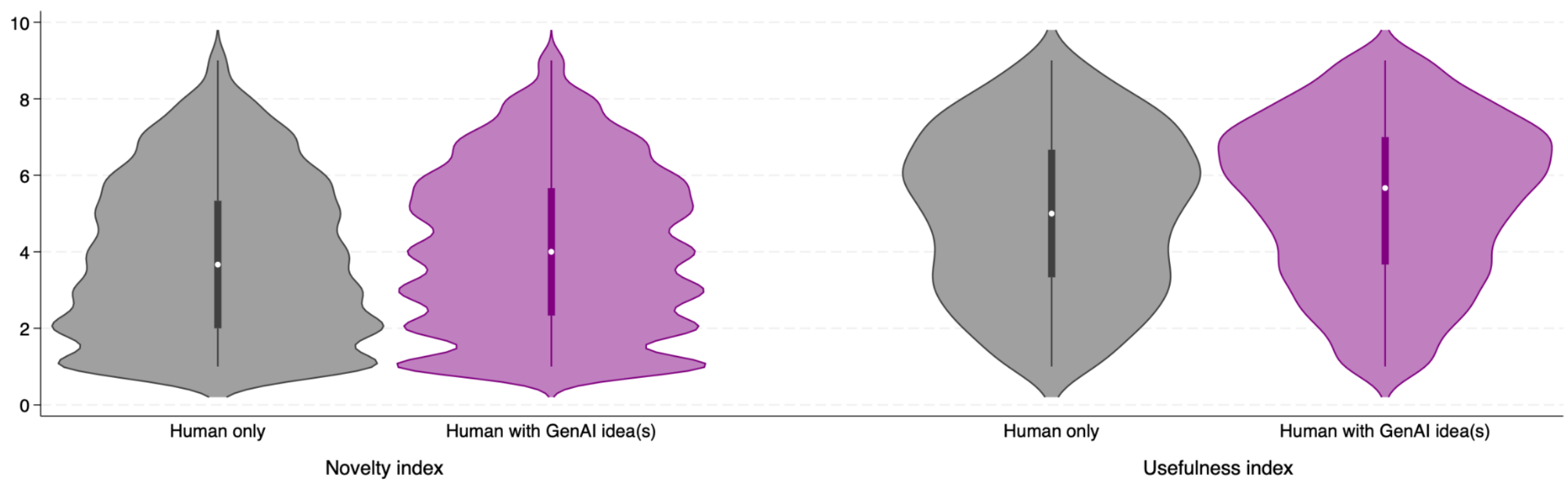

Supplementary Figure 2. Violin plot of conditions

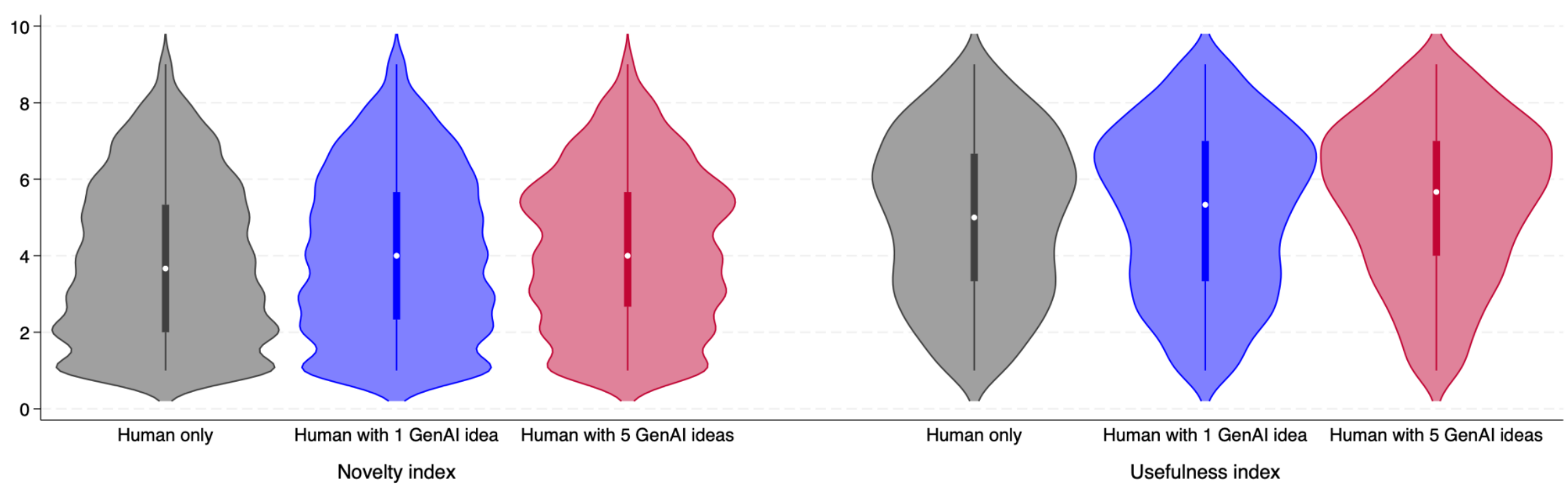

Supplementary Figure 3. Remainder of emotion outcomes

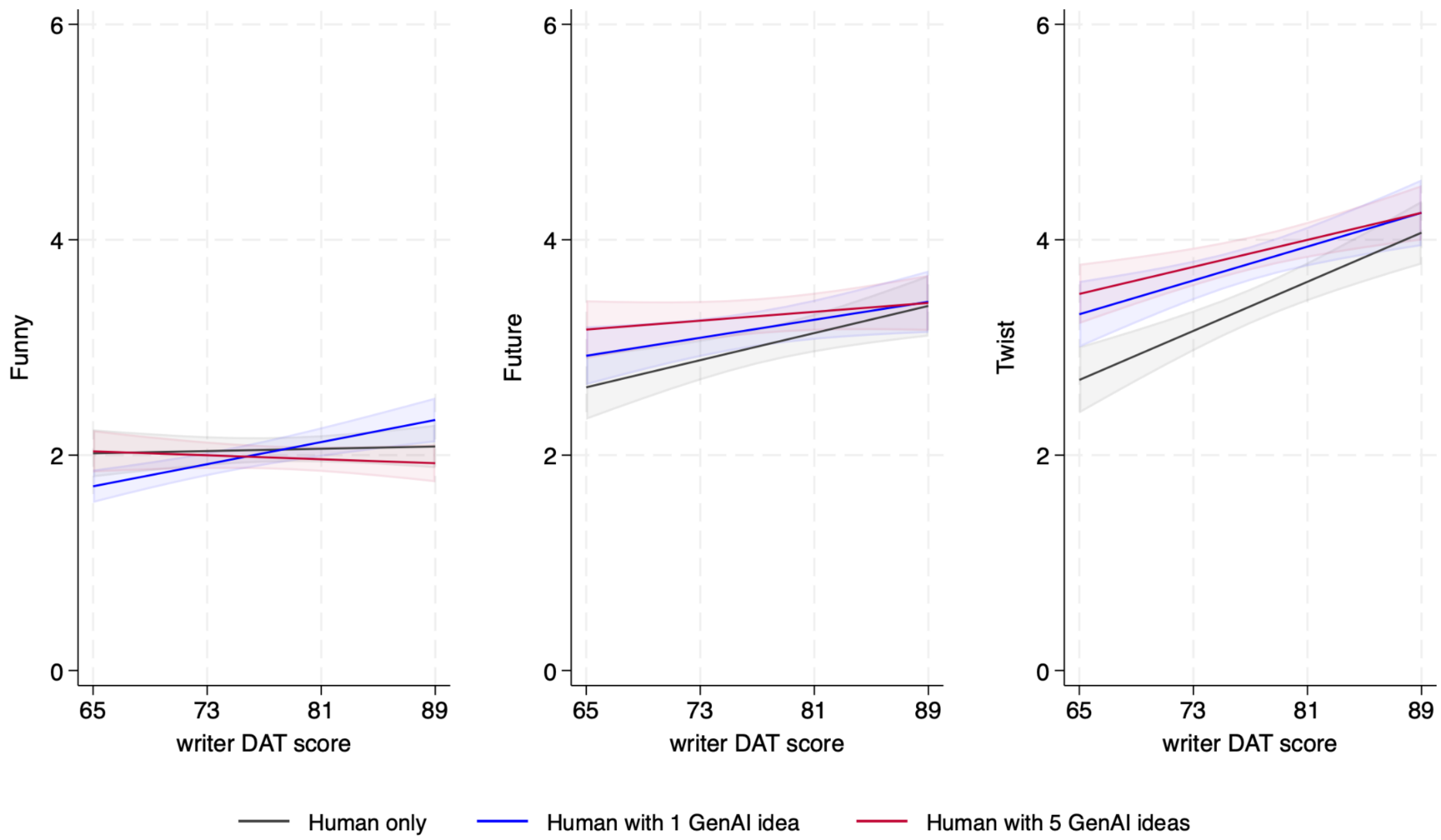

# Section 3. Similarity to AI ideas of *Human Only* Participants and AI Idea Participants Who Did Not Use AI

One concern is that participants in the *Human only* condition reported that they did not use AI, but did in fact do so. We conduct the following analysis, which suggests this is not the case. For the *Human only* condition and those in the GenAI idea conditions that did not opt to use GenAI, we provide a “simulated” AI story idea. We do this by randomly selecting one of the GenAI ideas that were generated for that topic and allocating it to participants who could not or did not access a GenAI idea. Then we look at the distribution of cosine similarity between a participant’s story and their GenAI story idea. We compare three groups: participants in the *Human only* condition who were randomly assigned a simulated GenAI idea for this analysis, participants in the GenAI idea conditions who did not use GenAI who were randomly assigned a simulated GenAI idea for this analysis, and participants in the GenAI idea conditions who used the GenAI idea during the study, comparing their final story to the first available GenAI idea.

A comparison of the distribution shows that the mode and range of the first two groups are more similar—stories of *Human only* participants look more like participants who chose not to generate a GenAI story. Those two distributions are less similar to their assigned GenAI ideas than the third group. Summary statistics of the three groups reflect this comparison as well.

## Supplementary Figure 4. Distribution of Story–AI idea similarity

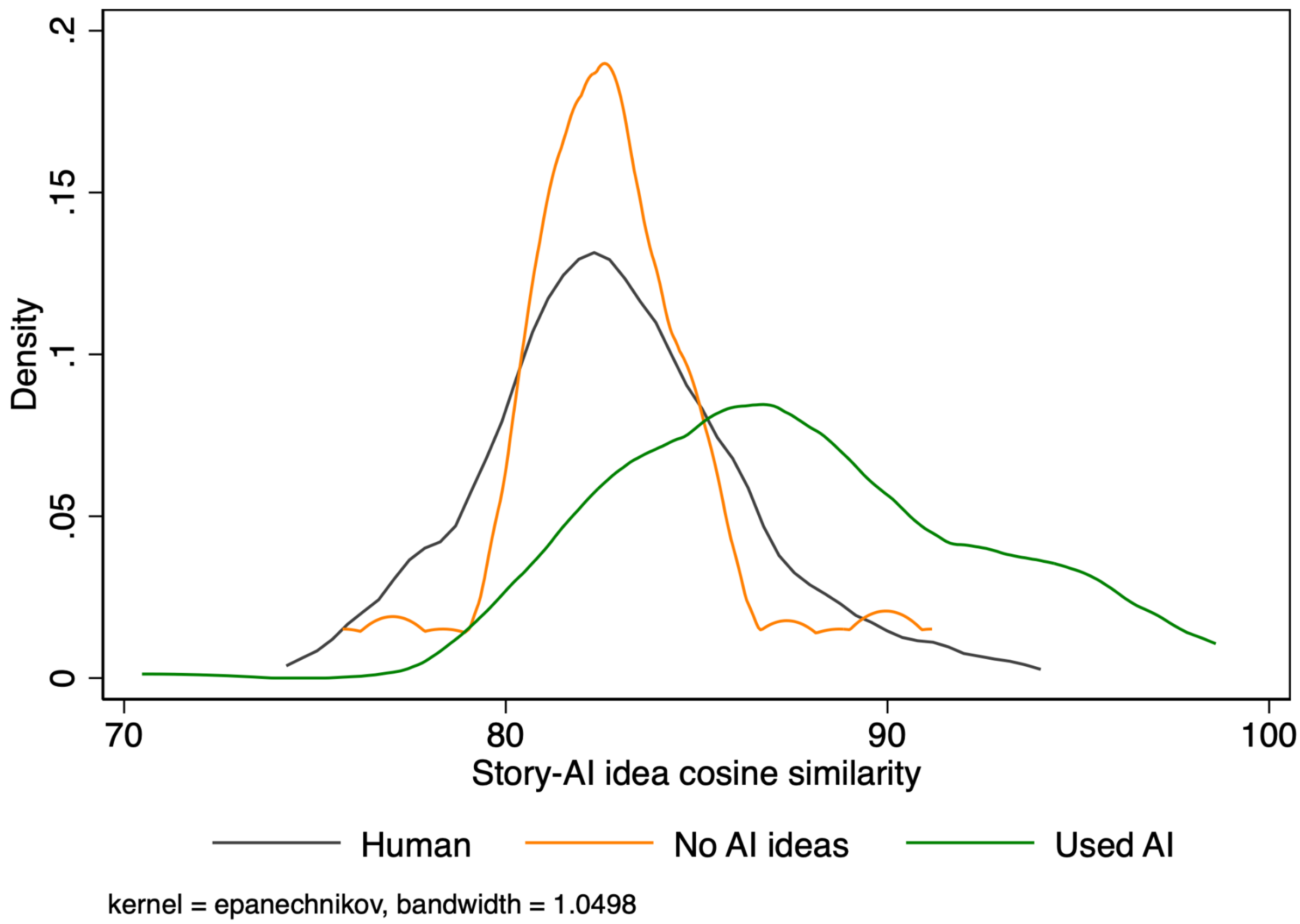

## Supplementary Table 18. Summary statistics for Story–AI similarity

| | Count | Mean | S.D. | 25th pctile | 50th pctile | 75th pctile |
|---|---|---|---|---|---|---|
| Human | 95 | 82.85 | 3.34 | 80.91 | 82.44 | 84.83 |
| No AI ideas | 23 | 82.89 | 3.10 | 81.45 | 82.21 | 84.15 |
| Used AI | 175 | 87.59 | 4.77 | 84.24 | 87.28 | 90.69 |

# Section 4. Pre-registered analysis (AsPredicted #136723)

# CONFIDENTIAL - FOR PEER-REVIEW ONLY

## Generative Artificial Intelligence and Creative Production (#136723)

**Created:** 06/26/2023 02:06 AM (PT)

---

This is an anonymized copy (without author names) of the pre-registration. It was created by the author(s) to use during peer-review.
A non-anonymized version (containing author names) should be made available by the authors when the work it supports is made public.

---

**1) Have any data been collected for this study already?**
No, no data have been collected for this study yet.

**2) What's the main question being asked or hypothesis being tested in this study?**
How does the availability of generative AI to assist with a creative task (i.e. writing a short story) affect the (self-)evaluation of the creative output by creators and third-party evaluators?

**3) Describe the key dependent variable(s) specifying how they will be measured.**
We are interested in assessing the following dependent variables, which will be measured for both creators and third-party evaluators:
- Novelty: an index of three questions on a scale of 1 to 9
- Usefulness: an index of three questions on a scale of 1 to 9
We will create aggregate indices for these measures, as well as look at their individual components.

**4) How many and which conditions will participants be assigned to?**
There are three conditions in the study:
- "Human only" condition where the project creator does not get any AI assistance
- "Hybrid" condition where the creator has the opportunity to access one short prompt for a story idea from OpenAI's ChatGPT API
- "Hybrid+" condition where the creator has the opportunity to access up to five short prompts for a story idea from OpenAI's ChatGPT API

**5) Specify exactly which analyses you will conduct to examine the main question/hypothesis.**
We will run OLS regressions predicting novelty and usefulness by condition, and run these regressions for both creators and evaluators.

We will run robustness tests for each of these, which will include different econometric specifications and variants (e.g., sub-items, discretization) of the outcome measures.

Note: While the Hybrid and Hybrid+ conditions technically differ in their capabilities (the latter allows for more AI prompts), we plan to combine the two conditions into one joint condition for our main analysis if the main outcome variables in those two conditions are not statistically significant from each other. (We will still report the existence of all three conditions and a results breakdown by all three conditions in the appendix.)

**6) Describe exactly how outliers will be defined and handled, and your precise rule(s) for excluding observations.**
We will exclude all participants that did not finish the study for analysis purposes. We will also drop respondents in the "Human only" condition that acknowledged that they used generative AI to assist with their responses.

**7) How many observations will be collected or what will determine sample size? No need to justify decision, but be precise about exactly how the number will be determined.**
For each condition, we will collect n=100 creators per condition who complete the study, for a total of n=300 creators across the three conditions. Then, we will collect n=600 third-party evaluators (each of which evaluates six stories drawn at random from the creator conditions).

**8) Anything else you would like to pre-register? (e.g., secondary analyses, variables collected for exploratory purposes, unusual analyses planned?)**
We have additional exploratory outcome variables about each story (e.g. enjoyment of the story, how well written, funny or boring the story is, etc.) on a scale from 1 to 9, for which we will study treatment effects similar to our main analysis.

We have collected a number of variables that we will use to look at heterogeneous effects including the creativity of the respondent (through a creativity task prior to the writing task and through self-ratings), their prior experience with generative AI technologies, and demographic information (e.g. gender, education, employment and income).

We will also consider non-linear relationships, by discretizing the outcomes and running linear probability models.

# Section 5. Writer study screenshots

# Study Overview

**Overview:** This study will consist of two parts and a short follow-up survey. In some parts, you will be asked understanding questions. You must answer these understanding questions correctly in order to proceed to complete the study.

**Payment:** For completing this study, you are guaranteed to receive a £3.00 within 48 hours. In addition, one part of the two parts will be randomly selected as the part-that-counts. Any amount (if any) you earn in the part-that-counts will be distributed to you as a bonus payment after 4-6 weeks.

**Understanding Question:** Which of the following statements is true?

- ○ For completing this study, I will receive £3 within 48 hours, but I do NOT have a chance of receiving any additional bonus payment.
- ○ For completing this study, I will receive £3 within 48 hours, and I will also receive the amount I earn in the part-that-counts as additional bonus payment
- ○ For completing this study, I will receive £3 within 48 hours, and I will also receive the total amount I earn across all parts as additional bonus payment.

Next

# Part 1

## Instructions

Please enter 10 words that are as **different** from each other as possible, in all meanings and uses of the words.

## Rules

- Only **single words** in English.
- Only **nouns** (e.g., things, objects, concepts).
- **No proper nouns** (e.g., no specific people or places).
- **No specialised vocabulary** (e.g., no technical terms).
- Think of the words **on your own** (e.g., do not just look at objects in your surroundings).

## Enter words

1.
2.
3.
4.
5.
6.
7.
8.

Condition: *Human only*

# Part 2

## Instructions

We would like you to write a story about **an adventure in the jungle**. You can write about anything you like. The story must be **exactly eight sentences long** and it needs to be written in English and appropriate for a **teenage and young adult audience** (approximately 15 to 24 years of age).

Please write your story (**exactly 8 sentences**) here:

Current sentence count: **0**

You need to write 8 sentences to enable the Next button.

Sentences end with a full stop (.) a question mark (?) or an exclamation mark (!) Next

Condition: *Human with 1 AI idea* (before story has been generated)

# Part 2

## Instructions

We would like you to write a story about **an adventure in the jungle**. You can write about anything you like. The story must be **exactly eight sentences long** and it needs to be written in English and appropriate for a **teenage and young adult audience** (approximately 15 to 24 years of age).

In order to assist you, we have provided you **access to AI assistance** that, if you wish, will come up with a starting point for your story by clicking on **"Generate Story Idea"**. The response from the AI assistant will be created in real time by a sophisticated AI algorithm. The response will be added in grey text below. **You are free to use or disregard any element of the AI assistant's story idea, or start over with your own idea**.

Generate Story Idea...

Please write your story (**exactly 8 sentences**) here:

Current sentence count: **0**

You need to write 8 sentences to enable the Next button.

Sentences end with a full stop (.) a question mark (?) or an exclamation mark (!) Next

Condition: *Human with 1 AI idea* (after story has been generated)

# Part 2

## Instructions

We would like you to write a story about **an adventure in the jungle**. You can write about anything you like. The story must be **exactly eight sentences long** and it needs to be written in English and appropriate for a **teenage and young adult audience** (approximately 15 to 24 years of age).

In order to assist you, we have provided you **access to AI assistance** that, if you wish, will come up with a starting point for your story by clicking on **"Generate Story Idea"**. The response from the AI assistant will be created in real time by a sophisticated AI algorithm. The response will be added in grey text below. **You are free to use or disregard any element of the AI assistant's story idea, or start over with your own idea**.

Three friends embark on a thrilling adventure in the unexplored depths of the Amazon rainforest in search of a legendary ancient city. Battling unpredictable weather, dangerous wildlife, and treacherous terrains, they uncover hidden tribal mysteries and ancient secrets. Their friendship and courage are tested to the limits as they face life-threatening challenges, leading to a climax that changes their lives forever.

Please write your story (**exactly 8 sentences**) here:

Current sentence count: **0**

You need to write 8 sentences to enable the Next button.

Sentences end with a full stop (.) a question mark (?) or an exclamation mark (!) Next

Condition: *Human with 5 AI ideas* (before story has been generated)

# Part 2

## Instructions

We would like you to write a story about **an adventure in the jungle**. You can write about anything you like. The story must be **exactly eight sentences long** and it needs to be written in English and appropriate for a **teenage and young adult audience** (approximately 15 to 24 years of age).

In order to assist you, we have provided you **access to AI assistance** that, if you wish, will come up with a starting point for your story by clicking on **"Generate Story Idea"**. The response from the AI assistant will be created in real time by a sophisticated AI algorithm. You may select each of the 5 tabs below, and click the "Generate Story Idea" button. The response will be added to that tab.

Here is a guide to the status of the AI request in each tab:

- Free You have not generated a story idea in this tab. Click to start generating a new story.
- Used The AI has provided you with a suggested starting point for your story. Click to view the AI's suggestion.
- Error Sometimes the AI is busy and the request fails. Click to request a new story.

In each Used tab, the AI assistant will come up with another story idea that will be added to the list of starting points for your story. **You are free to use or disregard any element of the AI assistant's story idea, or start over with your own idea.**

Idea 1 Free | Idea 2 Free | Idea 3 Free | Idea 4 Free | Idea 5 Free

Generate Story Idea...

Please write your story (**exactly 8 sentences**) here:

Current sentence count: **0**

You need to write 8 sentences to enable the Next button.

Sentences end with a full stop (.) a question mark (?) or an exclamation mark (!) Next

Condition: *Human with 5 AI ideas* (<u>after</u> story has been generated)

# Part 2

## Instructions

We would like you to write a story about **an adventure in the jungle**. You can write about anything you like. The story must be **exactly eight sentences long** and it needs to be written in English and appropriate for a **teenage and young adult audience** (approximately 15 to 24 years of age).

In order to assist you, we have provided you **access to AI assistance** that, if you wish, will come up with a starting point for your story by clicking on **"Generate Story Idea"**. The response from the AI assistant will be created in real time by a sophisticated AI algorithm. You may select each of the 5 tabs below, and click the "Generate Story Idea" button. The response will be added to that tab.

Here is a guide to the status of the AI request in each tab:

- Free You have not generated a story idea in this tab. Click to start generating a new story.
- Used The AI has provided you with a suggested starting point for your story. Click to view the AI's suggestion.
- Error Sometimes the AI is busy and the request fails. Click to request a new story.

In each Used tab, the AI assistant will come up with another story idea that will be added to the list of starting points for your story. **You are free to use or disregard any element of the AI assistant's story idea, or start over with your own idea.**

Idea 1 Used | Idea 2 Used | Idea 3 Free | Idea 4 Free | Idea 5 Free

While on a research expedition in the Amazon, a young scientist, Ava, discovers an ancient, mystical artifact in a hidden temple. The artifact's removal awakens a protective spirit, and Ava must navigate treacherous terrain, wild beasts, and magical obstacles to return it and calm the enraged spirit. Along her journey, she learns important lessons about respect for nature and ancient cultures, eventually succeeding in her quest and forging a profound connection with the Amazon.

Please write your story (**exactly 8 sentences**) here:

Current sentence count: **0**

You need to write 8 sentences to enable the Next button.

Sentences end with a full stop (.) a question mark (?) or an exclamation mark (!) Next

# Follow-up survey

We have a few questions about your experience today. You will need to complete all of the questions in order to receive your payment

Next

Only shown in condition: *Human only*

# Follow-up survey

**Please tell us whether you used ChatGPT or a similar generative AI tool to inspire your story?** (Please answer truthfully: your truthful answer will help us with our research. Your answer will NOT affect your payment.)

○ Yes ○ No

Next

Only shown in condition: *Human only* and answered “yes” on previous page

# Follow-up survey

Here is the story you have submitted:

........

To what extent do you think did the AI generated assistance affect the story you have submitted?

(Please answer truthfully: your truthful answer will help us with our research. Your answer will NOT affect your payment.)

| | Not at all<br>1 | 2 | 3 | 4 | 5 | 6 | 7 | 8 | Extremely<br>9 |
|---|---|---|---|---|---|---|---|---|---|
| **Please indicate to what extent the AI generated assistance affect the story you have submitted** | ○ | ○ | ○ | ○ | ○ | ○ | ○ | ○ | ○ |

You will need to rate the AI generated idea on order to enable the Next button

Next

# Follow-up survey

Here is the story you have submitted:

........

Please indicate you how much you agree with the following statements on the following scale: 1=Not at all, 5=Moderately, 9=Extremely:

| | Not at all<br>1 | 2 | 3 | 4 | 5 | 6 | 7 | 8 | Extremely<br>9 |
|---|---|---|---|---|---|---|---|---|---|
| **This story has a surprising twist.** | ○ | ○ | ○ | ○ | ○ | ○ | ○ | ○ | ○ |
| **This story has changed what I expect of future stories I will read.** | ○ | ○ | ○ | ○ | ○ | ○ | ○ | ○ | ○ |
| **This story is funny.** | ○ | ○ | ○ | ○ | ○ | ○ | ○ | ○ | ○ |
| **This story is boring.** | ○ | ○ | ○ | ○ | ○ | ○ | ○ | ○ | ○ |
| **This story is well written.** | ○ | ○ | ○ | ○ | ○ | ○ | ○ | ○ | ○ |
| **I enjoyed writing this story.** | ○ | ○ | ○ | ○ | ○ | ○ | ○ | ○ | ○ |

Next

# Follow-up survey

Here is the story you have submitted:

........

| | Not at all 1 | 2 | 3 | 4 | 5 | 6 | 7 | 8 | Extremely 9 |
|---|---|---|---|---|---|---|---|---|---|
| **To what extent do you think your story reflects your own ideas?** | ○ | ○ | ○ | ○ | ○ | ○ | ○ | ○ | ○ |

Next

# Follow-up survey

Here is the story you have submitted:

........

| | Not at all<br>1 | 2 | 3 | 4 | 5 | 6 | 7 | 8 | Extremely<br>9 |
|---|---|---|---|---|---|---|---|---|---|
| **How novel do you think your story is?** | ○ | ○ | ○ | ○ | ○ | ○ | ○ | ○ | ○ |
| **How original do you think your story is?** | ○ | ○ | ○ | ○ | ○ | ○ | ○ | ○ | ○ |
| **How rare (e.g. unusual) do you think your story is?** | ○ | ○ | ○ | ○ | ○ | ○ | ○ | ○ | ○ |

Next

# Follow-up survey

Here is the story you have submitted:

........

| | Not at all 1 | 2 | 3 | 4 | 5 | 6 | 7 | 8 | Extremely 9 |
|---|---|---|---|---|---|---|---|---|---|
| **How appropriate do you think is your story for the intended audience?** | ○ | ○ | ○ | ○ | ○ | ○ | ○ | ○ | ○ |
| **How feasible to do you think is your story to be developed into a complete book?** | ○ | ○ | ○ | ○ | ○ | ○ | ○ | ○ | ○ |
| **How likely do you think would it be that your story is turned into a complete book if a publisher read it and hired a professional author to expand on the idea?** | ○ | ○ | ○ | ○ | ○ | ○ | ○ | ○ | ○ |

Next

Only shown in condition: *Human with 1 AI idea*

# Follow-up survey

Here is the story you have submitted:

........

**To what extent do you think did this specific AI generated story idea affect the story you have submitted?**

(Please answer truthfully: your truthful answer will help us with our research. Your answer will NOT affect your payment.)

Three friends embark on a thrilling adventure in the unexplored depths of the Amazon rainforest in search of a legendary ancient city. Battling unpredictable weather, dangerous wildlife, and treacherous terrains, they uncover hidden tribal mysteries and ancient secrets. Their friendship and courage are tested to the limits as they face life-threatening challenges, leading to a climax that changes their lives forever.

| | Not at all 1 | 2 | 3 | 4 | 5 | 6 | 7 | 8 | Extremely 9 |
|---|---|---|---|---|---|---|---|---|---|
| **Please indicate to what extent this specific AI generated idea affected the story you submitted** | ○ | ○ | ○ | ○ | ○ | ○ | ○ | ○ | ○ |

You will need to rate the AI generated idea on order to enable the Next button

Next

Only shown in condition: *Human with 5 AI ideas* (before rating any story)

# Follow-up survey

Here is the story you have submitted:

........

**To what extent do you think did this specific AI generated story idea affect the story you have submitted?**

Please respond to this question for each tab with a response that has the Rate me icon

(Please answer truthfully: your truthful answer will help us with our research. Your answer will NOT affect your payment.)

Idea 1 Rate me | Idea 2 Rate me | Idea 3 Unused | Idea 4 Unused | Idea 5 Unused

When siblings Ryan and Emily discover an ancient map in their grandpa's attic, they become captivated by the promise of unknown treasure and launch an unprecedented journey into the dense, dangerous jungle. The siblings overcome deadly obstacles, ally with indigenous tribes, and unearth secrets of their family's explorative past. Their adventurous spirit leads them to an unimaginable treasure, providing not only wealth but also an understanding of their family's courageous legacy.

| | Not at all 1 | 2 | 3 | 4 | 5 | 6 | 7 | 8 | Extremely 9 |
|---|---|---|---|---|---|---|---|---|---|
| **Please indicate to what extent this specific AI generated idea affected the story you submitted** | ○ | ○ | ○ | ○ | ○ | ○ | ○ | ○ | ○ |

You will need to rate all of the AI generated ideas on order to enable the Next button

Next

Only shown in condition: *Human with 5 AI ideas* (after rating a story)

# Follow-up survey

Here is the story you have submitted:

........

**To what extent do you think did this specific AI generated story idea affect the story you have submitted?**

Please respond to this question for each tab with a response that has the Rate me icon

(Please answer truthfully: your truthful answer will help us with our research. Your answer will NOT affect your payment.)

Idea 1 ✓ Rated | Idea 2 Rate me | Idea 3 Unused | Idea 4 Unused | Idea 5 Unused

While on a research expedition in the Amazon, a young scientist, Ava, discovers an ancient, mystical artifact in a hidden temple. The artifact's removal awakens a protective spirit, and Ava must navigate treacherous terrain, wild beasts, and magical obstacles to return it and calm the enraged spirit. Along her journey, she learns important lessons about respect for nature and ancient cultures, eventually succeeding in her quest and forging a profound connection with the Amazon.

| | Not at all 1 | 2 | 3 | 4 | 5 | 6 | 7 | 8 | Extremely 9 |
|---|---|---|---|---|---|---|---|---|---|
| **Please indicate to what extent this specific AI generated idea affected the story you submitted** | ○ | ○ | ○ | ○ | ○ | ○ | ○ | ○ | ○ |

You will need to rate all of the AI generated ideas on order to enable the Next button

Next

# Follow-up survey

Here is the story you have submitted:

........

In your own words, who do you identify as having provided the original spark and idea for this story?

Next

# Follow-up survey

**If this story were published and sold tomorrow, how much of the story's profit do you believe should belong to you versus other entities (<u>such as prior books, stories, or AI tools</u>) that may have provided the starting point for your story?**

(Please answer truthfully: your truthful answer will help us with our research. Your answer will NOT affect your payment.)

**Please indicate the percentage of the story's profit that you believe you should receive:**

**50%**

0% 25% 50% 75% 100%

Next

# Follow-up survey

| | Not at all 1 | 2 | 3 | 4 | 5 | 6 | 7 | 8 | Extremely 9 |
|---|---|---|---|---|---|---|---|---|---|
| **How creative do you consider yourself?** | ○ | ○ | ○ | ○ | ○ | ○ | ○ | ○ | ○ |
| **How much creativity is required in your job?** | ○ | ○ | ○ | ○ | ○ | ○ | ○ | ○ | ○ |
| **How comfortable are you with new technologies?** | ○ | ○ | ○ | ○ | ○ | ○ | ○ | ○ | ○ |
| **How much (if at all) have you previously engaged with AI or similar technologies?** | ○ | ○ | ○ | ○ | ○ | ○ | ○ | ○ | ○ |

**Have you used any of the following AI tools in the past?** (Check all that apply)

- ☐ None
- ☐ ChatGPT
- ☐ Dall-E
- ☐ OpenAI's playground (e.g. DaVinci, Currie, Ada)
- ☐ Stable Diffusion
- ☐ NightCafe
- ☐ Jasper
- ☐ Microsoft Bing Chat
- ☐ Google Bard
- ☐ You.com
- ☐ Midjourney
- ☐ Other

Ai Tools Other Name

[ ]

**Have you used any of the following categories of AI tools in the past?** (Check all that apply)

- ☐ None
- ☐ Text
- ☐ Image
- ☐ Audio
- ☐ Music
- ☐ Video

Next

# Demographics

What gender do you identify with?

- ◯ Female
- ◯ Male
- ◯ Prefer not to say
- ◯ Other (please specify below)

Other gender

What is your current age? (enter a number of years)

What is your highest level of education?

- ◯ Less than A levels
- ◯ Vocational training
- ◯ A levels
- ◯ Undergraduate degree
- ◯ Postgraduate Master's degree
- ◯ Professional degree (e.g. MBA, JD)
- ◯ Doctorate

What is your current employment status?

- ◯ Employed full time
- ◯ Employed part time
- ◯ Unemployed looking for work
- ◯ Unemployed not looking for work
- ◯ Retired
- ◯ Student
- ◯ Disabled

What is your current job title?

What is your current annual income?

- ◯ Less than £10,000
- ◯ £10,000-£24,999
- ◯ £25,000-£49,999
- ◯ £50,000-£74,999
- ◯ £75,000-£99,999
- ◯ £100,000-£124,999
- ◯ £125,000-£149,999
- ◯ More than £150,00

Do you have any additional comments about this survey?

Next

Section 6. Evaluator study screenshot

# Study Overview

**Overview:** This study will consist of two main parts and a short follow-up survey. In each of the two main parts, you will be asked to read 6 short (eight-sentence) stories and answer a series of questions about them.

**Payment:** For completing this study, you are guaranteed to receive a £3 within 48 hours.

**Please answer all questions carefully and honestly.**

**Understanding Question:** Which of the following statements is true?

- ○ For completing this study, I will receive £3 within 48 hours plus an unspecified bonus payment.
- ○ For completing this study, I will receive £3 within 48 hours.
- ○ For completing this study, I will not receive any payment for this study.

Next

# Part 1: Additional instructions

In Part 1 of the study, we will show you **6 different, short stories**.

Each story will be shown on a different page, and each story is approximately eight sentences long, written in English, and intended for a **teenage and young adult audience** (approximately 15 to 24 years of age).

For each story, we ask that you to read it carefully and answer a series of questions about it.

The 6 short stories are about three different topics:

- 2 short stories are about an adventure on the open seas.
- 2 short stories are about an adventure in the jungle.
- 2 short stories are about an adventure on a different planet.

Stories that are about the same topic will be shown one after the another. The order in which you will see the stories and topics is as shown in the list above.

Next

# Story (1 of 6)

**Instructions:** After reading this story, we will ask you a series of questions. The story will remain on the screen after you press continue. Note the story is intended for a **teenage and young adult audience** (approximately 15 to 24 years of age).

Here is the story you are reviewing:

**Topic:** Write a short story about an adventure on a different planet

Callum and his sister Beth had travelled to many planets with their mother as part of an exploratory mission. There were twelve other teenagers on the ship and they had grown up with each other, experiencing the thrill and dangers of space travel together. As they got older their parents allowed them to take part more in their work. The first time they actually got to do this was on a previously unknown planet called Xephyr . The aliens living on Xephyr were curious to meet the crew and their families and Callum and Beth found that their teenage equivalent on Xephyr were not that different from themselves and they enjoyed making new friends . However, the next uncharted planet was very different...

Please read the story carefully and then advance to the next page.

(You will need to wait at least 10s before you can click the Next button)

Next

# Story (1 of 6) Review

Here is the story you are reviewing:

**Topic:** Write a short story about an adventure on a different planet

Callum and his sister Beth had travelled to many planets with their mother as part of an exploratory mission. There were twelve other teenagers on the ship and they had grown up with each other, experiencing the thrill and dangers of space travel together. As they got older their parents allowed them to take part more in their work. The first time they actually got to do this was on a previously unknown planet called Xephyr . The aliens living on Xephyr were curious to meet the crew and their families and Callum and Beth found that their teenage equivalent on Xephyr were not that different from themselves and they enjoyed making new friends . However, the next uncharted planet was very different...

Please indicate you how much you agree with the following statements on the following scale: 1=Not at all, 5=Moderately, 9=Extremely:

| | Not at all<br>1 | 2 | 3 | 4 | 5 | 6 | 7 | 8 | Extremely<br>9 |
|---|---|---|---|---|---|---|---|---|---|
| **This story is funny.** | ○ | ○ | ○ | ○ | ○ | ○ | ○ | ○ | ○ |
| **This story has changed what I expect of future stories I will read.** | ○ | ○ | ○ | ○ | ○ | ○ | ○ | ○ | ○ |
| **This story is boring.** | ○ | ○ | ○ | ○ | ○ | ○ | ○ | ○ | ○ |
| **This story is well written.** | ○ | ○ | ○ | ○ | ○ | ○ | ○ | ○ | ○ |
| **I enjoyed reading this story.** | ○ | ○ | ○ | ○ | ○ | ○ | ○ | ○ | ○ |
| **This story has a surprising twist.** | ○ | ○ | ○ | ○ | ○ | ○ | ○ | ○ | ○ |

Next

# Story (1 of 6) Review

Here is the story you are reviewing:

**Topic:** Write a short story about an adventure on a different planet

Callum and his sister Beth had travelled to many planets with their mother as part of an exploratory mission. There were twelve other teenagers on the ship and they had grown up with each other, experiencing the thrill and dangers of space travel together. As they got older their parents allowed them to take part more in their work. The first time they actually got to do this was on a previously unknown planet called Xephyr . The aliens living on Xephyr were curious to meet the crew and their families and Callum and Beth found that their teenage equivalent on Xephyr were not that different from themselves and they enjoyed making new friends . However, the next uncharted planet was very different...

| | Not at all<br>1 | 2 | 3 | 4 | 5 | 6 | 7 | 8 | Extremely<br>9 |
|---|---|---|---|---|---|---|---|---|---|
| **How novel do you think the story is?** | ○ | ○ | ○ | ○ | ○ | ○ | ○ | ○ | ○ |
| **How original do you think the story is?** | ○ | ○ | ○ | ○ | ○ | ○ | ○ | ○ | ○ |
| **How rare (e.g. unusual) do you think the story is?** | ○ | ○ | ○ | ○ | ○ | ○ | ○ | ○ | ○ |

Next

# Story (1 of 6) Review

Here is the story you are reviewing:

**Topic:** Write a short story about an adventure on a different planet

Callum and his sister Beth had travelled to many planets with their mother as part of an exploratory mission. There were twelve other teenagers on the ship and they had grown up with each other, experiencing the thrill and dangers of space travel together. As they got older their parents allowed them to take part more in their work. The first time they actually got to do this was on a previously unknown planet called Xephyr . The aliens living on Xephyr were curious to meet the crew and their families and Callum and Beth found that their teenage equivalent on Xephyr were not that different from themselves and they enjoyed making new friends . However, the next uncharted planet was very different...

| | Not at all 1 | 2 | 3 | 4 | 5 | 6 | 7 | 8 | Extremely 9 |
|---|---|---|---|---|---|---|---|---|---|
| **How appropriate do you think the story is for the intended audience?** | ○ | ○ | ○ | ○ | ○ | ○ | ○ | ○ | ○ |
| **How feasible to do you think the story is to be developed into a complete book?** | ○ | ○ | ○ | ○ | ○ | ○ | ○ | ○ | ○ |
| **How likely do you think would it be that the story is turned into a complete book if a publisher read it and hired a professional author to expand on the idea?** | ○ | ○ | ○ | ○ | ○ | ○ | ○ | ○ | ○ |

Next

# Part 2: Additional instructions

Thank you for completing Part 1.

In Part 2 of the study, we will show you again the **same 6 short stories.**

This time you will be asked **different questions** than before.

After that, you will answer a short follow-up survey and demographic questions.

Next

# Story (1 of 6)

Here is the story you are reviewing:

**Topic:** Write a short story about an adventure on a different planet

Callum and his sister Beth had travelled to many planets with their mother as part of an exploratory mission. There were twelve other teenagers on the ship and they had grown up with each other, experiencing the thrill and dangers of space travel together. As they got older their parents allowed them to take part more in their work. The first time they actually got to do this was on a previously unknown planet called Xephyr . The aliens living on Xephyr were curious to meet the crew and their families and Callum and Beth found that their teenage equivalent on Xephyr were not that different from themselves and they enjoyed making new friends . However, the next uncharted planet was very different...

Please indicate the extent (if any) to which you think this story was based on inputs from an AI tool (e.g. ChatGPT or similar generative AI tool):

**0%**

0% 25% 50% 75% 100%

Next

Example of a story from the *Human Only* condition: the writer said they did not use an outside AI tool

# Story (1 of 6) Review

Here is the story you are reviewing:

**Topic:** Write a short story about an adventure on a different planet

Many years into the future, a new planet was discovered. The planet was very far away, and not many people knew of its existence. When the first colonists started to arrive there, it was still very wild and dangerous. Despite the danger, many people were drawn to there in pursuit of wealth and adventure. Many others were trying to escape into the unknown wilderness. The rapid advancement of technology had improved many aspects of society on paper, but left some people disillusioned with their lives. These people wanted to return to a more primitive and meaningful way of life. The newly discovered remote planet provided many with the opportunity.

The author of this story was **NOT** provided with **access to AI assistance**. To the best of our knowledge, the author **did not** consult any other AI tool. In short, the author **wrote this story without any input from AI**.

| | Not at all 1 | 2 | 3 | 4 | 5 | 6 | 7 | 8 | Extremely 9 |
|---|---|---|---|---|---|---|---|---|---|
| **To what extent do you think the story reflects the author's own ideas?** | ○ | ○ | ○ | ○ | ○ | ○ | ○ | ○ | ○ |
| **To what extent does the author have an "ownership" claim to the final story?** | ○ | ○ | ○ | ○ | ○ | ○ | ○ | ○ | ○ |

Next

Example of a story from the *Human Only* condition: the writer said they did use an outside AI tool

# Story (3 of 6) Review

Here is the story you are reviewing:

**Topic:** Write a short story about an adventure on the open seas

........

The author of this story was **NOT** provided with **access to AI assistance**. But the author indicated that **they consulted another AI tool**. In short, the author may have **written the story with input from AI**.

| | Not at all 1 | 2 | 3 | 4 | 5 | 6 | 7 | 8 | Extremely 9 |
|---|---|---|---|---|---|---|---|---|---|
| **To what extent do you think the story reflects the author's own ideas?** | ○ | ○ | ○ | ○ | ○ | ○ | ○ | ○ | ○ |
| **To what extent does the author have an "ownership" claim to the final story?** | ○ | ○ | ○ | ○ | ○ | ○ | ○ | ○ | ○ |

Next

Example of a story from the *Human with 1 or 5 AI idea(s)* condition but the author did not access the GenAI idea

# Story (2 of 6) Review

Here is the story you are reviewing:

**Topic:** Write a short story about an adventure on a different planet

........

The author of this story was provided with **access to AI assistance**, which could come up with a starting point for their story. The author **did NOT** choose to make use of the AI assistance. To the best of our knowledge, the author **did not** consult any other AI tool. In short, the author **wrote this story without any input from AI**.

| | Not at all 1 | 2 | 3 | 4 | 5 | 6 | 7 | 8 | Extremely 9 |
|---|---|---|---|---|---|---|---|---|---|
| **To what extent do you think the story reflects the author's own ideas?** | ○ | ○ | ○ | ○ | ○ | ○ | ○ | ○ | ○ |
| **To what extent does the author have an "ownership" claim to the final story?** | ○ | ○ | ○ | ○ | ○ | ○ | ○ | ○ | ○ |

Next

Example of a story from the *Human with 1 GenAI idea* condition: the writer requested a GenAI idea

## Story (3 of 6) Review

Here is the story you are reviewing:

**Topic:** Write a short story about an adventure in the jungle

Within the heart of an ancient abandoned jungle, an explorer called Robert was exploring this jungle which once had a many tribes on it. While exploring this jungle he found what was left of some of these tribes, abandoned huts, old fire places, and bones. He didn't know what could of caused all of the tribes to die out. As he was walking, Robert felt a strange presence in the jungle, he saw some footsteps going deep in the jungle.
He followed the footsteps until he found a temple which was build inside a mountain, He walked inside of the temple he saw 3 paths going into different directions, each of the paths were lit up by torches, he knew what that meant. Robert took the path that went straight, but said that these paths went extremely deep into the mountain, until he heard some sounds, that were of the remaining people of this jungle, he thought to himself that they might be the people that killed all the other tribes. Robert watched from away as the people were having their meal, when he stumbled and they heard him and turned around, He saw what looked to be the leader of the tribe walking towards him. Robert didn't try to run away but question the people to understand what was happening, the leader said that they didn't know about the other tribes being dead, and they were very friendly with him, they invited him to have his meal with them, and offered him to stay the night, he accepted the offer, he went to sleep, and after that the leader of the tribe killed him in his sleep.

The author of this story was provided with **access to AI assistance**, which could come up with a starting point for their story. The author **chose to make use** of the AI assistance and the AI assistance provided the following prompt(s) to the author:

**AI suggestion 1**
While exploring a mysterious ancient temple deep in the heart of the jungle, a group of friends accidentally awakens a long-forgotten entity that guards the temple. Now they must race against time to decrypt hidden clues and solve cryptic puzzles to deactivate the ancient curse before it destroys everything in its path. Friendships are tested, secrets are revealed, and through their daring adventure, the group discovers hidden strengths within themselves that give them the courage to save their lives and secure the temple's treasures.

In short, the author may have **written the story with input from AI**.

| | Not at all 1 | 2 | 3 | 4 | 5 | 6 | 7 | 8 | Extremely 9 |
|---|---|---|---|---|---|---|---|---|---|
| **To what extent do you think the story reflects the author's own ideas?** | ○ | ○ | ○ | ○ | ○ | ○ | ○ | ○ | ○ |
| **To what extent does the author have an "ownership" claim to the final story?** | ○ | ○ | ○ | ○ | ○ | ○ | ○ | ○ | ○ |

If this story were published and sold tomorrow, how much of the story's profit do you believe should belong to **the author** versus **the creators of the generative AI tool** that may have provided the starting point for the story?

**0%**

0% 25% 50% 75% 100%

Next

Example of a story from the *Human with 5 GenAI ideas* condition: the writer requested five GenAI ideas

## Story (1 of 6) Review

Here is the story you are reviewing:

**Topic:** Write a short story about an adventure on a different planet

Callum and his sister Beth had travelled to many planets with their mother as part of an exploratory mission. There were twelve other teenagers on the ship and they had grown up with each other, experiencing the thrill and dangers of space travel together. As they got older their parents allowed them to take part more in their work. The first time they actually got to do this was on a previously unknown planet called Xephyr . The aliens living on Xephyr were curious to meet the crew and their families and Callum and Beth found that their teenage equivalent on Xephyr were not that different from themselves and they enjoyed making new friends . However, the next uncharted planet was very different...

The author of this story was provided with **access to AI assistance**, which could come up with a starting point for their story. The author **chose to make use** of the AI assistance and the AI assistance provided the following prompt(s) to the author:

**AI suggestion 1**
Three friends, Abigail, Elliot, and Maddox, accidentally stumble upon an intergalactic portal, landing on the mysterious Planet Zephyr. Battling harsh alien terrain, inhospitable creatures and limited oxygen supplies, they embark on a daring mission to locate a legendary power source reportedly capable of returning them back to Earth. The trio's bravery, intelligence, and unyielding spirit are put to the ultimate test, in a high-stakes, high-adrenaline off-world adventure unlike anything they'd ever imagined.

**AI suggestion 2**
When a group of young astronauts land on the uncharted planet Xerxes V, they discover a vibrant ecosystem filled with sentient plant life. Challenged by the complex society of these plants and a rapidly dwindling oxygen supply, the group must harness their courage, intellect and adaptability to understand the alien civilization and work with them to survive. The unexpected unity of human and plant life on Xerxes V results in an astonishing adventure depicting interstellar diplomacy and the overwhelming power of cooperation.

**AI suggestion 3**
Three earthlings inadvertently stumble upon a portal leading to an exotic yet perilous planet named Zephyros. Their brave adventures are filled with encounters of strange, alien species, near-death experiences due to the planet's unpredictable weather, and the resolution of an age-old Zephyri conflict. Eventually, they return to Earth with tales of their remarkable interstellar adventures and hitherto unknown insights about the cosmos.

**AI suggestion 4**
In quest of finding signs of life beyond Earth, a group of astronauts embark on a thrilling journey to Prolion-6, an unexplored planet in a distant galaxy. They encounter a complex ecosystem of intelligent, peaceful alien species and discover ancient technologies far advanced than humanity's. Their adventure reveals a profound interstellar culture with cosmic potential, prompting both awe and existential thoughts.

**AI suggestion 5**
Three human explorers journey to an uncharted planet, Epsilon Prime, in search of precious minerals. They encounter alien life that communicates through color-changing patterns on their skin, causing initial conflict due to misunderstood intentions. Through cooperation, both parties learn to understand each other, leading to a newfound interspecies alliance and the successful completion of the explorer's mission.

In short, the author may have **written the story with input from AI**.

| | Not at all 1 | 2 | 3 | 4 | 5 | 6 | 7 | 8 | Extremely 9 |
|---|---|---|---|---|---|---|---|---|---|
| **To what extent do you think the story reflects the author's own ideas?** | ○ | ○ | ○ | ○ | ○ | ○ | ○ | ○ | ○ |
| **To what extent does the author have an "ownership" claim to the final story?** | ○ | ○ | ○ | ○ | ○ | ○ | ○ | ○ | ○ |

If this story were published and sold tomorrow, how much of the story's profit do you believe should belong to **the author** versus **the creators of the generative AI tool** that may have provided the starting point for the story?

**0%**

0% 25% 50% 75% 100%

Next

# AI Overview

Please indicate you how much you agree with the following statements on the following scale: 1=Not at all, 5=Moderately, 9=Extremely:

| | Not at all 1 | 2 | 3 | 4 | 5 | 6 | 7 | 8 | Extremely 9 |
|---|---|---|---|---|---|---|---|---|---|
| **If a human creator (author) uses AI in part of the writing of a story, the AI-generated content should be accessible alongside the final story.** | ○ | ○ | ○ | ○ | ○ | ○ | ○ | ○ | ○ |
| **It is ethically acceptable to use AI to come up with an initial idea for a story.** | ○ | ○ | ○ | ○ | ○ | ○ | ○ | ○ | ○ |
| **If AI is used in any part of the writing of a story, the creators of the content on which the AI output was based on should be compensated.** | ○ | ○ | ○ | ○ | ○ | ○ | ○ | ○ | ○ |
| **If AI is used in any part of the writing of a story, the final story no longer counts as a "creative act".** | ○ | ○ | ○ | ○ | ○ | ○ | ○ | ○ | ○ |
| **Relying on the use of AI to write a new story is unethical.** | ○ | ○ | ○ | ○ | ○ | ○ | ○ | ○ | ○ |
| **It is ethically acceptable to use AI to write and publicly disseminate an entire story without acknowledging the use of AI.** | ○ | ○ | ○ | ○ | ○ | ○ | ○ | ○ | ○ |

Next

# Follow-up survey

| | Not at all<br>1 | 2 | 3 | 4 | 5 | 6 | 7 | 8 | Extremely<br>9 |
|---|---|---|---|---|---|---|---|---|---|
| **How creative do you consider yourself?** | ○ | ○ | ○ | ○ | ○ | ○ | ○ | ○ | ○ |
| **How much creativity is required in your job?** | ○ | ○ | ○ | ○ | ○ | ○ | ○ | ○ | ○ |
| **How comfortable are you with new technologies?** | ○ | ○ | ○ | ○ | ○ | ○ | ○ | ○ | ○ |
| **How much (if at all) have you previously engaged with AI or similar technologies?** | ○ | ○ | ○ | ○ | ○ | ○ | ○ | ○ | ○ |

**Have you used any of the following AI tools in the past?** (Check all that apply)

- ☐ None
- ☐ ChatGPT
- ☐ Dall-E
- ☐ OpenAI's playground (e.g. DaVinci, Currie, Ada)
- ☐ Stable Diffusion
- ☐ NightCafe
- ☐ Jasper
- ☐ Microsoft Bing Chat
- ☐ Google Bard
- ☐ You.com
- ☐ Midjourney
- ☐ Other

Ai Tools Other Name

**Have you used any of the following categories of AI tools in the past?** (Check all that apply)

- ☐ None
- ☐ Text
- ☐ Image
- ☐ Audio
- ☐ Music
- ☐ Video

Next

# Demographics

What gender do you identify with?

- ◯ Female
- ◯ Male
- ◯ Prefer not to say
- ◯ Other (please specify below)

Other gender

[ ]

What is your current age? (enter a number of years)

[ ]

What is your highest level of education?

- ◯ Less than A levels
- ◯ Vocational training
- ◯ A levels
- ◯ Undergraduate degree
- ◯ Postgraduate Master's degree
- ◯ Professional degree (e.g. MBA, JD)
- ◯ Doctorate

What is your current employment status?

- ◯ Employed full time
- ◯ Employed part time
- ◯ Unemployed looking for work
- ◯ Unemployed not looking for work
- ◯ Retired
- ◯ Student
- ◯ Disabled

What is your current job title?

[ ]

What is your current annual income?

- ◯ Less than £10,000
- ◯ £10,000-£24,999
- ◯ £25,000-£49,999
- ◯ £50,000-£74,999
- ◯ £75,000-£99,999
- ◯ £100,000-£124,999
- ◯ £125,000-£149,999
- ◯ More than £150,00

Do you have any additional comments about this survey?

[ ]

Next